\documentclass[fleqn,usenatbib]{mnras}

\usepackage{newtxtext,newtxmath}

\usepackage[T1]{fontenc}
\usepackage{multirow}
\usepackage{multicol}
\DeclareRobustCommand{\VAN}[3]{#2}
\let\VANthebibliography\thebibliography
\def\thebibliography{\DeclareRobustCommand{\VAN}[3]{##3}\VANthebibliography}
\newcommand{\erf}{\ensuremath{\operatorname{erf}}}

\usepackage{graphicx}	
\usepackage{amsmath}	

\usepackage{amssymb}	

\newcommand{\liana}[1]{\textbf{\textcolor{blue}{Liana: #1}}}

\title{Exploring binary black hole mergers and host galaxies with {\sc Shark} and COMPAS}

\author[Rauf \textit{et al.}]{
Liana Rauf,$^{1}$\thanks{E-mail: l.rauf@uq.net.au}
Cullan Howlett,$^{1}$
Tamara~M.~Davis,$^{1}$
Claudia D.~P. Lagos,$^{2,3,4}$
\\
$^{1}$School of Mathematics and Physics, The University of Queensland, Brisbane, QLD 4072, Australia\\
$^{2}$International Centre for Radio Astronomy Research (ICRAR), M468, University of Western Australia, 35 Stirling Hwy, Crawley, WA 6009, Australia\\
$^{3}$ARC Centre of Excellence for All Sky Astrophysics in 3 Dimensions (ASTRO 3D)\\
$^{4}$Cosmic Dawn Center (DAWN), Denmark.
}

\date{Accepted XXX. Received YYY; in original form ZZZ}

\pubyear{2023}

\begin{document}
\label{firstpage}
\pagerange{\pageref{firstpage}--\pageref{lastpage}}
\maketitle

\begin{abstract}
We explore the connection between the gravitational wave (GW) merger rates of stellar-mass binary black holes (BBH) and galaxy properties. We do this by generating populations of stars using the binary population synthesis code COMPAS and evolving them in galaxies from the semi-analytic galaxy formation model {\sc Shark}, to determine the number of mergers occurring in each simulation time-step. We find that metal-rich and massive galaxies with star formation rate (SFR) greater than $1M_{\odot}/ \rm yr$ are 10 times more likely to have GW events compared to younger, less massive and metal poor galaxies. Our simulation with the default input parameters predicts a higher local merger rate density compared to the third GW transient catalogue (GWTC-3) prediction from LIGO, VIRGO and KAGRA, due to short coalescence times, low metallicities and a high SFR at low redshift in the simulation, which produces more BBHs that merge within the age of the Universe compared to observations. We identify alternate remnant mass models that more accurately reproduce the volumetric rate and provide updated fits to the merger rate as a function of redshift. We then investigate the relative fraction of GW events in our simulation that are in observable host galaxies from upcoming galaxy surveys, determining which of those are ideal for tracing host galaxies with high merger rates. The implications of this work can be utilised for constraining stellar evolution models, better informing follow-up programs, and placing informative priors on host galaxies when measuring cosmological parameters such as the Hubble constant.
\end{abstract}




\section{Introduction}
The direct detection of GWs emitted from a binary black hole (BBH) coalescence in 2015 was a significant moment in history.  Since then, BBHs, along with other compact objects, have been detected frequently by the LIGO/VIRGO collaboration during their observing runs \citep{2021PhRvX..11b1053A}. Combining the data from GWTC-1, GWTC-2.1 and GWTC-3, there is a total of 90 GW event candidates, including 85 BBHs, 2 binary neutron stars (BNSs), and 3 black-hole-neutron-stars (BHNSs) with $p_{\rm astro}>0.5$ (\citealt{ligo_scientific_collaboration_2022_7249086}). \par 
The cross-correlation between spatial distributions of GW sources and galaxies will play a significant role in constraining the relationship between luminosity distance and redshift, thus constraining the Hubble constant and other cosmological parameters \citep{2020ApJ...902...79B,mukherjee2022cross}. GWs could hold the key to resolving the tension between early-type and late-type measurements of $H_0$. An independent measurement of $H_0$ from multi-messenger astronomy could confirm if the current tension is a statistical anomaly or evidence of new physics that cannot be explained via the Lambda Cold Dark Matter ($\Lambda$CDM) model of cosmology \citep{2020ApJ...900L..33P}. 

GWs also offer insight into binary formation mechanisms and can be used to inform our understanding of stellar evolution from the measurements of the source astrophysical parameters. On a larger scale, GWs sources provide a new tool to study the Universe and its cosmological parameters. However, these applications of GWs are limited unless we know exactly where these events are located in the sky \textit{or} which galaxies are more likely to produce them. Due to a lack of understanding of various dynamical processes, there is an uncertainty associated with the mass spectrum of BHs. Several mechanisms have been proposed in the literature, and studies such as \citet{spera2015mass} and \citet{fryer2012compact} have derived theoretical models for the BH mass as a function of metallicity and main-sequence mass based on the physical models in population synthesis codes. Understanding how the merger rate is affected by these stellar evolution prescriptions can have a huge impact on gravitational wave astronomy. In the future, these predicted merger rates can be validated against the detected mergers in order to constrain stellar evolution models for compact binary systems. Specifically, GW events and their host galaxy's history can help constrain compact binary merger formation channels \citep{2017ApJ...846...82Z, 2020ApJ...905...21A}. Finally, localisation of GW sources and their electromagnetic (EM) counterparts is currently not very accurate \citep{cao2018host}, with uncertainty on the sky localisation still on the order of tens of square degrees \citep{2020LRR....23....3A}. If the host galaxy properties of these GW events can be understood, these galaxies can be tracked down more efficiently, since the search for their sources can be constrained. Thus, statistical simulations of GW hosts will enable better follow-up of GW triggers by putting priors on the types of galaxies to host GWs. \par 
There have been studies that have looked at the origins of the first BNS merger event, GW170817 \citep{2018A&A...615A..91B, 2023NatAs...7..444S}. Other recent works such as \citet{artale2019host, artale2019mass}, \citet{2020ApJ...905...21A}, \citet{2021MNRAS.502.4877S, 2022MNRAS.516.3297S} and \citet{2021ApJ...907..110B} have further investigated the merger rates for compact objects as a function of global host galaxy properties. \citet{artale2019host, artale2019mass} investigate the merger rate per galaxy over cosmic time for BBHs, binary neutron stars (BNS) and black hole-neutron star binaries (BHNS) using simulated data from EAGLE \citep{schaye2015eagle}. They found a strong correlation between the merger rate and stellar mass. Overall, the merger rate is high for galaxies with high stellar mass, star formation rate (SFR) and metallicity. \cite{2020ApJ...905...21A} investigates only the merger rate history of BNSs. They found that there are two distinct trends depending on the coalescence time of the BNSs. BNSs with short coalescence times prefer to merge in galaxies with high SFR and BNSs with longer coalescence times prefer galaxies with high stellar mass at $z=0$. \cite{2021MNRAS.502.4877S} found that a low mass transfer efficiency and high common envelop ejection efficiency resulted in a low BBH merger rate history. Due to their mass, BBHs are more affected by dynamics and metallicity evolution than BNSs. Similar conclusions are made in \cite{2022MNRAS.516.3297S}, where they explore varying the scaling relations for galaxies such as the stellar mass function, star formation rate distribution and mass-metallicity and fundamental metallicity relation. The BBH merger rate is steeper using the mass-metallicity relation compared to the fundamental metallicity relation, as the latter has a shallow decrease of metallicity with redshift. \cite{2021ApJ...907..110B} assumes an isolated binary evolution scenario for modelling the formation of double compact objects (DCOs). They found that the formation efficiency, which is the number of DCOs formed per unit mass, increases towards low metallicities for BBHs. \par 
More recently, \citet{2022MNRAS.514.2716M} also used EAGLE combined with stellar evolution codes, BPASS and COSMIC to track the evolution of compact binaries from birth to coalescence. The majority of the binaries evolving in the EAGLE galaxies were born within 0.5 $\leq z \leq$ 3 and merge within 0 $\leq z \leq$ 1.They also found that 55-70\% of merging compact binaries are born in late-type galaxies. \citet{2022MNRAS.514.1315B} combines stellar models from BPASS with various cosmological simulations (Empirical, Millennium, EAGLE and Illustris-TNG) to estimate transient rates for supernovae, long gamma-ray bursts and GWs.\par 
In this paper, we combine a semi-analytic model for galaxy evolution and formation, called {\sc Shark} \citep{2018MNRAS.481.3573L}, with the population synthesis code, COMPAS \citep{2022ApJS..258...34R, stevenson2017formation, 2018MNRAS.481.4009V}, to establish a model for the merger rate of BBHs. \citet{2022arXiv220714126M} uses COMPAS with GALFORM \citep{cole2000hierarchical, lacey2016unified} to probe the nature of dark matter with BBH mergers. However, this is the first time {\sc Shark} has been used for modelling GW rates and galaxy properties. It is also the first time semi-analytic modelling has been utilised to investigate the galaxy photometry and its correlation with the GW rate. Other studies have used hydrodynamical simulations for generating their galaxy catalogues, with little to no analysis on the photometry. This is vital in constraining localisation of host galaxies in current and upcoming surveys such as DESI \citep{2021MNRAS.tmp..319R, 2019AJ....157..168D}, LSST \citep{2019BAAS...51c.363M}, 4MOST \citep{richard20194most} and WALLABY \citep{koribalski2020wallaby}. \par
Our approach follows the work of \citet{artale2019host}, but by extending their approach to focus on observable galaxy properties, rather than just intrinsic galaxy properties, we endeavour to answer some crucial questions: 
\begin{itemize}
    \item What is the relationship between the BBH merger rate and \textit{observed} host galaxy properties? 
    \item How do alterations to stellar evolution models and formation mechanisms affect our estimation of the merger rate? 
    \item What photometric selections maximise the number of observed host galaxies with GW events?  
\end{itemize} 
We develop a pipeline to model the relationship between the merger rate of BBHs and their host galaxy properties, using simulated data to investigate how host galaxy properties affect the GW rate of binary black holes. Designing and implementing a merger rate model requires the use of the Initial Mass Function (IMF; \citealt{salpeter1955luminosity, kroupa2001variation, chabrier2003galactic}), the remnant mass spectrum and various BBH properties. \par 
This paper is structured as follows. In Section \ref{Section 2} we discuss {\sc Shark}, the semi-analytic model utilised to obtain the host galaxy properties, specifically, how it is constructed and the outputs that are utilised for our pipeline. Similarly, in Section \ref{Section 3} we discuss COMPAS, the input distributions necessary to generate our synthetic population and the relevant outputs for our model. In Section \ref{Section 4} we employ a forward-modelling approach, where we populate a realistic galaxy simulation using a model for the entire evolution of binary star system; from star formation to collapse. This enabled us to create a comprehensive catalogue of GW rates and their host galaxy properties in the local Universe. We display our final results in Section \ref{Section 5} with their implications on the merger rate in galaxies and how they vary with respect to the choice of stellar evolution models. In Section \ref{Section 6}, we investigate current and upcoming surveys and their target selections to predict the number of GW host galaxies and total GW rate based on our model. Our conclusions are given in Section \ref{Section 7}.

\section{{\sc Shark}: Semi-analytic modelling of galaxy formation and evolution} \label{Section 2}

In this work, we produce a simulation of GWs by tying a binary population synthesis model to a simulated galaxy catalogue containing physical and photometric parameters. The galaxy simulation we use is {\sc Shark} \citep{2018MNRAS.481.3573L}, a semi-analytic model for galaxy formation written in C++, and designed to explore various physical processes and ways to model them\footnote{Code is freely available at \url{https://github.com/ICRAR/shark}}. Studies of galaxy evolution and formation must include realistic cosmological environments and effects. Hence, {\sc Shark} simulations use up-to-date parameters within a $\Lambda$CDM cosmology, fit to observations from the Planck Collaboration \citep{ade2016planck}, to define the background cosmological model of the simulated universe. The growth of large scale structures is dominated by dark matter (assumed to be driven by gravity). The {\sc Shark} model is run over a dark-matter-only, $N$-body simulation called SURFS \citep{elahi2018surfs}, which has a comoving cosmological volume (units given as $\rm cMpc^3$) of $(210h^{-1}\,\rm cMpc)^3$. The galaxies are evolved in dark matter halos and subhalos across 200 snapshots (sub-haloes start forming in snapshot 20 and the first galaxies are recorded in snapshot 30), spanning between $z=0-24$. This is done by fully modelling the time-evolution of processes that modify the mass, metallicity and structural properties of the baryonic components in the simulation by numerically solving a set of ordinary differential equations for mass, metals and angular momentum exchange between baryon reservoirs. The reservoirs are different types of gas (hot, cold, atomic, molecular and ionised)  associated with the galaxy and dark matter halo. The halos and galaxies are tracked and their global properties are calculated with VELOCIRAPTOR and TreeFROG \citep{elahi2019hunting, elahi2019climbing}. \par 
The {\sc Shark} outputs used in this work are: 
\begin{itemize}
\item Stellar mass ($M_{*}$; $M_\odot$)
\item Star Formation Rate (SFR; $M_\odot\mathrm{yr}^{-1}$)
\item Interstellar medium (ISM) gas metallicity ($Z_{\rm gas}$; dimensionless) - referred to as metallicity hereafter, and stellar metallicity ($Z_\star$)
\item Snapshot lookback time (Gyr)
\item The apparent magnitudes of galaxies at their observed redshift in the following wavebands: $u$ ($354$nm), $g$ ($475$nm), $r$ ($622$nm), $i$ ($763$nm), $z$ ($905$nm), $J$ ($1250$nm), $H$ ($1650$nm), $K$ ($2150$nm).
\end{itemize}
The {\sc Shark} outputs can be used to build lightcones, which shows the number counts and redshift distribution of galaxies in different wavebands \citep{2019MNRAS.489.4196L, 2020MNRAS.499.1948L}. This mock survey is created by applying the selection functions from a real survey, such as WALLABY, to our galaxy population in the simulation box. The 107 deg$^2$ lightcone used in this study was built with the code STINGRAY, a lightcone software originally developed by \citet{obreschkow2009virtual} and improved in \citet{Chauhan2019} to aid with the design and operation of telescopes such as the Square Kilometre Array.\footnote{Code is freely available at \url{https://github.com/obreschkow/stingray}} It is split into 64 subvolumes from which we utilised 5 of these subvolumes.\footnote{We use here the term ``subvolume'' but in practice, each of these 64 subvolumes is close to a random sample of the merger trees in the whole box, which are chosen in a way so that the use of a fraction of those does not lead to important cosmic variance. This also applies to the lightcones, where using a fraction of the 64 subvolumes does not introduce cosmic variance effects. Hence, using $5$ ``subvolumes'' gives us an effective survey area of $\approx 8.4$~deg$^2$.} The selection applied to this lightcone is $M_* > 10^5 M_\odot$ and a dummy magnitude cut of $\sim 32.5$, which is constructed by assuming a stellar mass to light ratio of $1$.

The key ingredients for modelling the BBH merger rate and host galaxy properties are the SFR and metallicity. The remainder of this section will compare the {\sc Shark} simulation outputs to current observations to affirm its validity.

\begin{figure}
    \centering
    \includegraphics[width=\columnwidth]{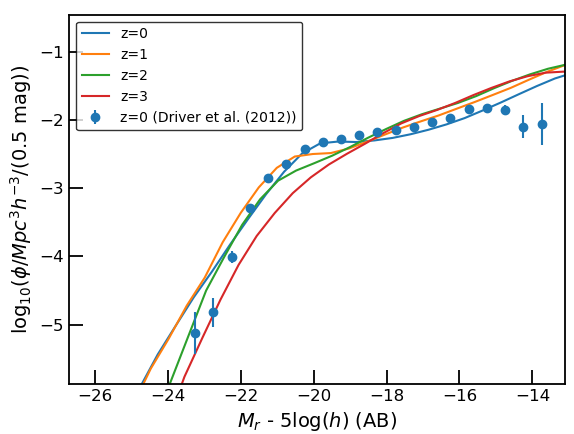}
    \caption{The rest-frame $r$-band luminosity function at various redshifts, measured from the {\sc Shark} galaxies in the simulation volume. The blue points are the observational measurements from \citet{driver2012galaxy} at $z=0$. Both {\sc Shark} and observational data are presented in bins of 0.5 mag.}
    \label{fig:lmf}
\end{figure}
Fig.~\ref{fig:lmf} shows the $r$-band galaxy luminosity function in {\sc Shark} at various redshifts, where $\phi$ is the number density of galaxies in each $r$-band magnitude bin. The observational points from the GAMA survey agree well with {\sc Shark}, except in the fainter bins  due to the GAMA survey being flux limited, and the corrections for this being more uncertain in the data. Similarly, at the brighter end, {\sc Shark} slightly overproduces the number density of bright galaxies, but cosmic variance plays a role here and is difficult to include in the observational error robustly. Nonetheless, the general agreement between the data and simulation predictions gives confidence that we can use {\sc Shark} luminosities in this work to make predictions for the real Universe. More thorough comparisons with multi-wavelength data of galaxies across cosmic time have been presented in previous {\sc Shark} papers \citep{2020MNRAS.497.3026B, 2018MNRAS.481.3573L, 2019MNRAS.489.4196L, 2020MNRAS.499.1948L}.\par 
\begin{figure}
    \centering
    \includegraphics[width=\columnwidth]{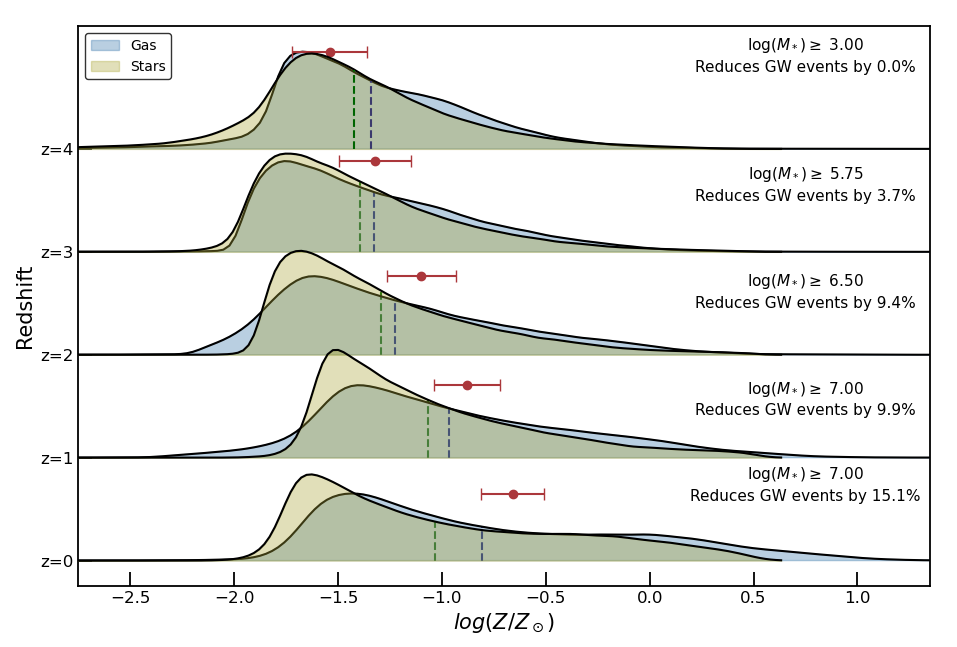}
    \caption{Stellar and gas metallicity distribution for {\sc Shark} galaxies. The y-axis is the snapshot redshift and at each redshift we apply stellar mass cuts to remove galaxies below the mass resolution limit. This reduces the total merger rate of the galaxies in the snapshots ($z=0,1,2,3,4$) by a small percentage, as given next to each histogram. The dashed vertical lines show the mean of each histogram, and the red points show observed gas metallicities fit from the metallicity evolution of damped Ly$\alpha$ systems \citep{rafelski2012metallicity}. We expect the dashed lines to be within the uncertainty range of the observational mean.}
    \label{fig:Z_dist}
\end{figure}
Fig.~\ref{fig:Z_dist} shows the stellar and ISM gas metallicity distributions in {\sc Shark}. The metallicity is calculated from the ratio of the mass in metals to the total mass and normalised to the solar metallicity, $Z_\odot = 0.012$. The distributions have a skewed log-normal shape, with a peak at low metallicities but broad distribution such that the mean metallicity is increasing with decreasing redshift. Overall, as stars evolve, the variance in metallicity increases with decreasing redshift. High gas and stellar metallicities  are more likely at low redshift, where previous episodes of star formation have enriched the gas. A consequence of this is that we would expect larger mass black holes to be formed in the earlier Universe, as mass loss from stellar winds becomes stronger with increasing metallicity \citep{fryer2012compact}. In this figure, we also include observational measurements of the mean metallicity of the neutral gas in damped Ly$\alpha$ systems from \cite{rafelski2012metallicity}, and see good agreement between these and the {\sc Shark} mean metallicities. However, in order to produce this plot, we have applied stellar mass cuts to the {\sc Shark} output to remove a small number (15\% of the galaxies at $z=0$ and less at higher redshift) of low-mass objects whose metallicities are less reliable due to the resolution limit of the SURFS simulation - in fact, \citet{2018MNRAS.481.3573L} suggest a reasonable resolution limit to be $M_{\star}=10^8\,\rm M_{\odot}$. Including these low-mass systems moves the mean metallicity in {\sc Shark} to lower than the observed values, but we note that these systems are at low enough mass that they do not contribute substantially to the formation rate of black holes --- we can still be confident that {\sc Shark} reproduces well the metallicities of galaxies that we expect to host the vast majority of BBHs at all redshifts we consider. In addition, \citet{2018MNRAS.481.3573L} presented a comparison between the predicted {\sc Shark} ISM gas and stellar metallicities scaling with stellar mass and a compilation of observations at $z\approx 0$, showing that within the uncertainties the model predictions agreed with observations.  \par 
\begin{figure}
    \centering
    \includegraphics[width=\columnwidth]{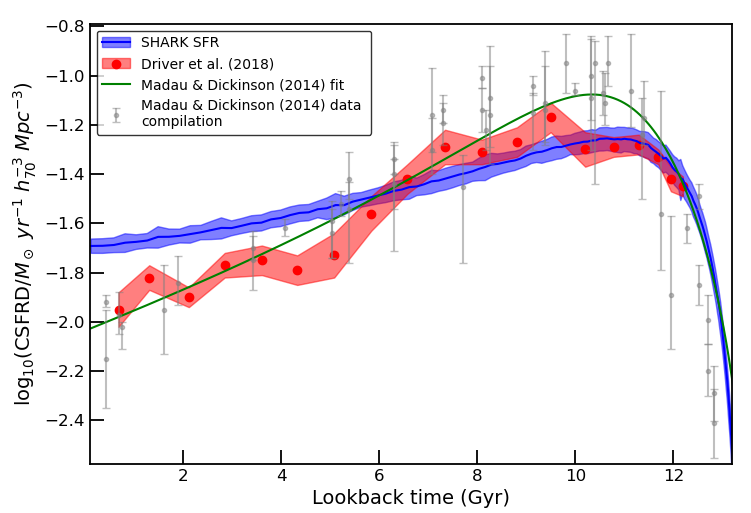}
    \caption{The cosmic SFR density in {\sc Shark} in blue and \citet{driver2018gama} in red, as a function of lookback time. The red shaded region is the uncertainty due to cosmic variance. The blue line is the average SFR density from all subvolumes in the {\sc Shark} simulation, and the uncertainty shaded in blue is the standard deviation. The grey points are the compilation of observational data presented in \citet{madau2014cosmic}, which themselves come from various studies referenced therein, and the green line is the best fitting function to the data.}
    \label{fig:CSFR}
\end{figure}
Finally, in Fig.~\ref{fig:CSFR}, we compare the {\sc Shark} SFR cosmic density to the observational data of \citet{madau2014cosmic} and \citet{driver2018gama}. {\sc Shark} and \citet{driver2018gama} agree well in the early Universe, where we observe the SFR being higher, and the gas metallicity lower, compared to the local Universe. Due to the low metallicity and high SFR, we expect the majority of BBHs to form in this period. However, they begin to diverge at a lookback time of 6 Gyr, reaching a maximum difference of a factor of $\sim 2$ at $z=0.0$. This over-prediction is due to more star-forming galaxies being quenched and transitioning to passive galaxies at a higher stellar mass in {\sc Shark} than observed, and an overproduction of stars at $z < 1$ due to star formation in the disk. These can be counteracted by increasing the strength of the active galactic nuclei (AGN) feedback. However, this causes tension in the low redshift, high mass end of the observational stellar mass function (SMF), which {\sc Shark} otherwise reproduces reasonably well. If the AGN feedback strength is increased, the {\sc Shark} predicted SMF falls under the observational SMF \citep{wright2017galaxy}. From previous works such as \cite{artale2019mass}, we expect the merger rate to have the strongest correlation with the stellar mass, so it is convenient that the accuracy of the {\sc Shark} SMF is prioritised. We discuss the consequences of this on our predictions for the GW rate in Section~\ref{Section 5}, but consider the star-formation rates from {\sc Shark} largely trustworthy given the good agreement at the times/redshifts most appropriate for our work. Also, \citet{2018MNRAS.481.3573L} presented a comparison with the observed SFR-stellar mass relation in the local Universe that agrees well with observations. In addition, \citet{2021MNRAS.500.2036K} show that {\sc Shark} produces specific SFR functions for galaxies in different bins of stellar mass in reasonable agreement with observations of the local Universe.

\section{COMPAS: Population synthesis tool for double compact objects} \label{Section 3} 

In addition to {\sc Shark}, we require a tool to populate our host galaxies with double compact objects. Therefore, we generate our population of binaries with a rapid Monte-Carlo population synthesis called Compact Object Mergers: Population and Statistics (COMPAS, \citealt{2022ApJS..258...34R, stevenson2017formation, 2018MNRAS.481.4009V}). In this section, we discuss the properties from COMPAS that are essential to our modelling. \par 
COMPAS was built to specifically study compact binaries that are sources of GWs. These binaries evolve through an isolated channel, where each pair of stars undergoes mass transfer and a common envelope phase. Using fitting formulae based on \citet{hurley2002evolution} and some user-defined initial conditions, COMPAS can infer the properties of the population generated at various evolutionary stages. Uncertainties in stellar evolution can also be explored, as COMPAS allows the model parameters to be easily changed in the code. In the literature, channels have been divided into two main categories; (a) evolution of isolated binaries and (b) dynamical formation in dense environments, such as globular clusters \citep{2021MNRAS.507.5224B,2021hgwa.bookE..16M,2022Galax..10...76S,2023MNRAS.520.5259A}. \citet{2021MNRAS.507.5224B} investigates the number of detections expected from each channel after varying the metallicity spread and spin parameters. The isolated channel is preferred for low-spin spin distribution and high metallicity spread, but excludes a purely isolated formation channel for GWTC-2. \citet{zevin2021one} also found that a mixture of isolated and dynamical channels is strongly preferred based on the GWTC-2 population of mergers, with a preference for the isolated channel. These studies rely on theoretical models and simplified assumptions, but there is evidence from observations that more than one formation channel contributes to the current observed population of GWs \citep{2016ApJ...819..108B,2020A&A...636A.104B,2020ApJ...899L...1Z,2022ApJ...940..171R}. Using COMPAS allows us to consider merger channel (a) only. \par 
In our work, the binary star primary masses are drawn from a Kroupa IMF \citep{kroupa2001variation}, although we discuss how changing this also changes our results later. Our fiducial analysis then draws the secondary mass from the mass ratio distribution given in \citet{sana2012binary}, but again we also explore an alternative flat distribution. We allow for chemically homogeneous evolution, using the ``pessimistic" model \citep{2022ApJS..258...34R}. \citet{2021MNRAS.505..663R} finds that this isolated formation channel produces up to 70\% of BBH mergers. The other input parameters for COMPAS are semi-major axis, period, eccentricity and metallicity. For these we adopt the default COMPAS distributions, or pass in data from {\sc Shark}. Table~\ref{tab:fid_case} summarises the distributions assumed in our fiducial simulations.

\begin{table}
    \centering
    \begin{tabular}{c|c}
       \hline
       \hline
       Distribution & Model  \\
       \hline
         IMF & \citet{kroupa2002initial}  \\
         Semi-major axis & Flat-in-log \\
         Orbital period & Flat-in-log \\
         Mass ratio & \citet{sana2012binary}  \\
         Eccentricity & \citet{sana2012binary}  \\
         Metallicity & {\sc Shark} \citep{2018MNRAS.481.3573L} \\
         Remnant mass & \citet{fryer2012compact} \\
       \hline
    \end{tabular}
    \caption{Summary of COMPAS input models for the ``fiducial" case.}
    \label{tab:fid_case}
\end{table}

\begin{figure}
    \centering
    \includegraphics[width=\columnwidth]{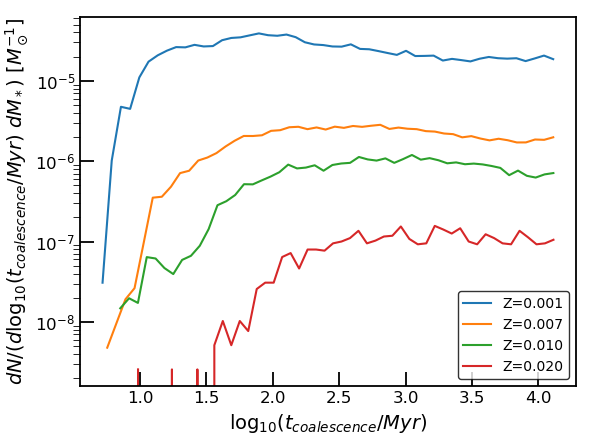}
    \includegraphics[width=\columnwidth]{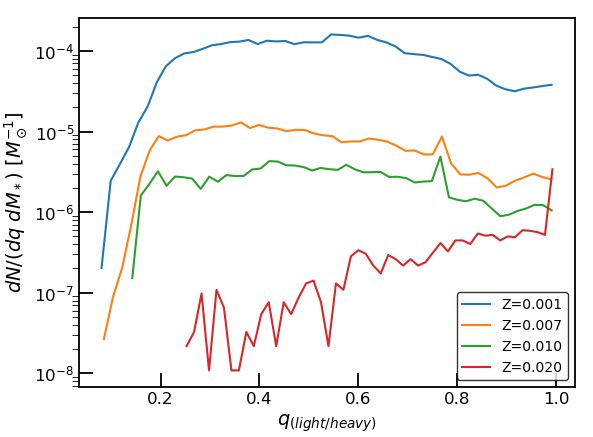}
    \caption{Coalescence time (\textit{top}) and mass ratio (\textit{bottom}) distributions for the BBHs produced in COMPAS for fixed metallicity bins, using the \citet{muller2016simple} remnant mass prescription. The y-axis is the number of binaries per unit star forming mass, per coalescence time or mass ratio bin.}
    \label{fig:COMPAS_dist}
\end{figure}

Fig.~\ref{fig:COMPAS_dist} shows our distributions of merging coalescence times (defined as the time between formation and coalescence), $t_{\mathrm{coalescence}}$, and black hole mass ratios, $q$, synthesised with COMPAS for binaries that manage to merge within the age of our simulated Universe. Low metallicity objects are more likely to form BBHs and merge within a Hubble time, hence both the coalescence time and mass ratio distributions are dominated by low metallicity systems. In addition, the shape of the distributions indicate that most of the compact objects from COMPAS have short coalescence times (typically on the order of tens to hundreds of Myr). The coalescence time distribution is qualitatively similar to the results of \cite{neijssel2019effect}. The mass ratio distribution tends to peak around $q=0.6$ at $Z=0.001$, with a gradual decay for $q > 0.6$. For $Z=0.007$ and $Z=0.010$ binaries with bigger mass differences (smaller mass ratios) are more common. For $Z=0.020$, both primary and secondary BBH masses must be relatively small due to the high metallicity.

\subsection{BBH Efficiency} 
\label{sec:feff}
There are some formation mechanisms that prohibit the formation of compact remnant objects. Therefore, only a small fraction of the COMPAS population will evolve into BBHs and merge. This is denoted as the BBH efficiency and can be roughly defined as, 
\begin{align} 
f_{\rm COMPAS} = \frac{N_{\rm BBH}}{N_{\rm total}},
\label{eq:f_eff}
\end{align}
where $N_{\rm BBH}$ is the number of BBHs and $N_{\rm total}$ is the total number of binary systems in our COMPAS run. Note that the stellar metallicity plays a significant role in determining the stellar evolution of the stars in binaries, and hence the BBH efficiency \citep{2018MNRAS.474.2959G,2019MNRAS.482.5012C,2019MNRAS.488.5300C,2022MNRAS.516.5737B}. 
We can consider this BBH efficiency factor as a rough scaling of the BBH merger rate and study how different choices for the COMPAS inputs will affect the overall merger rates, which we present later in this paper. For our fiducial case given in Table \ref{tab:fid_case}, $f_{\rm COMPAS} = 0.0194_{-0.0014}^{+0.0039}$. 

\begin{table}
    \centering
    \begin{tabular}{c|c|c}
        \hline
        \hline
        & Parameters & $f_{\rm COMPAS}$ \\
        \hline
        Case 1 & Mass ratio = Flat & $0.0208_{-0.0016}^{+0.0045}$ \vspace{1mm} \\
        Case 2 & Eccentricity = Flat & $0.0191_{-0.0013}^{+0.0039}$ \vspace{1mm} \\
        Case 3 & Semi-major axis = \citet{sana2012binary} & $0.0155_{-0.0013}^{+0.0022}$ \vspace{1mm} \\
        Case 4 & Remnant mass = \citet{hurley2000comprehensive} & $0.00202_{-0.0004}^{+0.0009}$ \vspace{1mm} \\
        Case 5 & Remnant mass = \citet{belczynski2008compact} & $0.0365_{-0.0028}^{+0.0031}$ \vspace{1mm} \\ 
        Case 6 & Remnant mass = \citet{muller2016simple} & $0.0027_{-0.0006}^{+0.0010}$ \vspace{1mm} \\
        Case 7 & Remnant mass = \citet{mandel2020simple} & $0.0026_{-0.0005}^{+0.0011}$ \vspace{1mm} \\
        Case 8 & Remnant mass = \citet{schneider2021pre} & $0.0021_{-0.0005}^{+0.0010}$ \\ 
        \hline
    \end{tabular}
    \caption{Summary of cases explored in this paper, where one parameter has been altered with respect to the fiducial model. The effective factor from COMPAS, $f_{\rm COMPAS}$, which is the fraction of BBHs formed in the COMPAS run, is also included. For $f_{\rm COMPAS}$ we show the average and range over 170 COMPAS runs (see \ref{method} for how we generate COMPAS runs for each {\sc Shark} snapshot).}
    \label{tab:cases}
\end{table}


\begin{figure}
    \centering
    \includegraphics[width=\columnwidth]{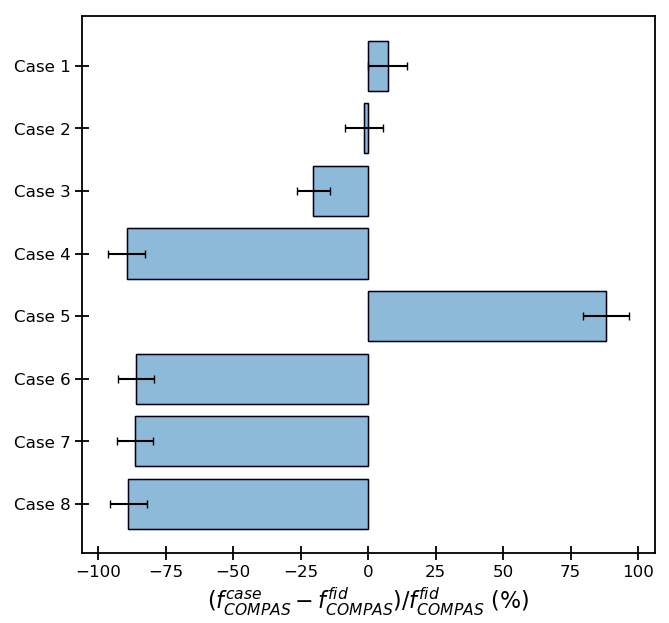}
    \caption{Percentage of BBHs formed in COMPAS relative to the fiducial case. $f_{\rm COMPAS}^{\rm fid}$ is the effective factor for our fiducial model and $f_{\rm COMPAS}^{\rm fid}$ is the effective factor for our cases.}
    \label{fig:effective_factor}
\end{figure}

In Table \ref{tab:cases}, we show the BBH efficiency using Eq.~\ref{eq:f_eff} for different COMPAS runs, where we vary only one input parameter from our fiducial case. The relative change in the BBH efficiency is shown in Fig.~\ref{fig:effective_factor}. The choice of stellar evolution models and the impact of this BBH fraction will be discussed in Section \ref{Section 5}. 

\section{Method: Derivation of BBH merger model} \label{Section 4} 

\begin{figure}
    \centering
    \includegraphics[width=\columnwidth]{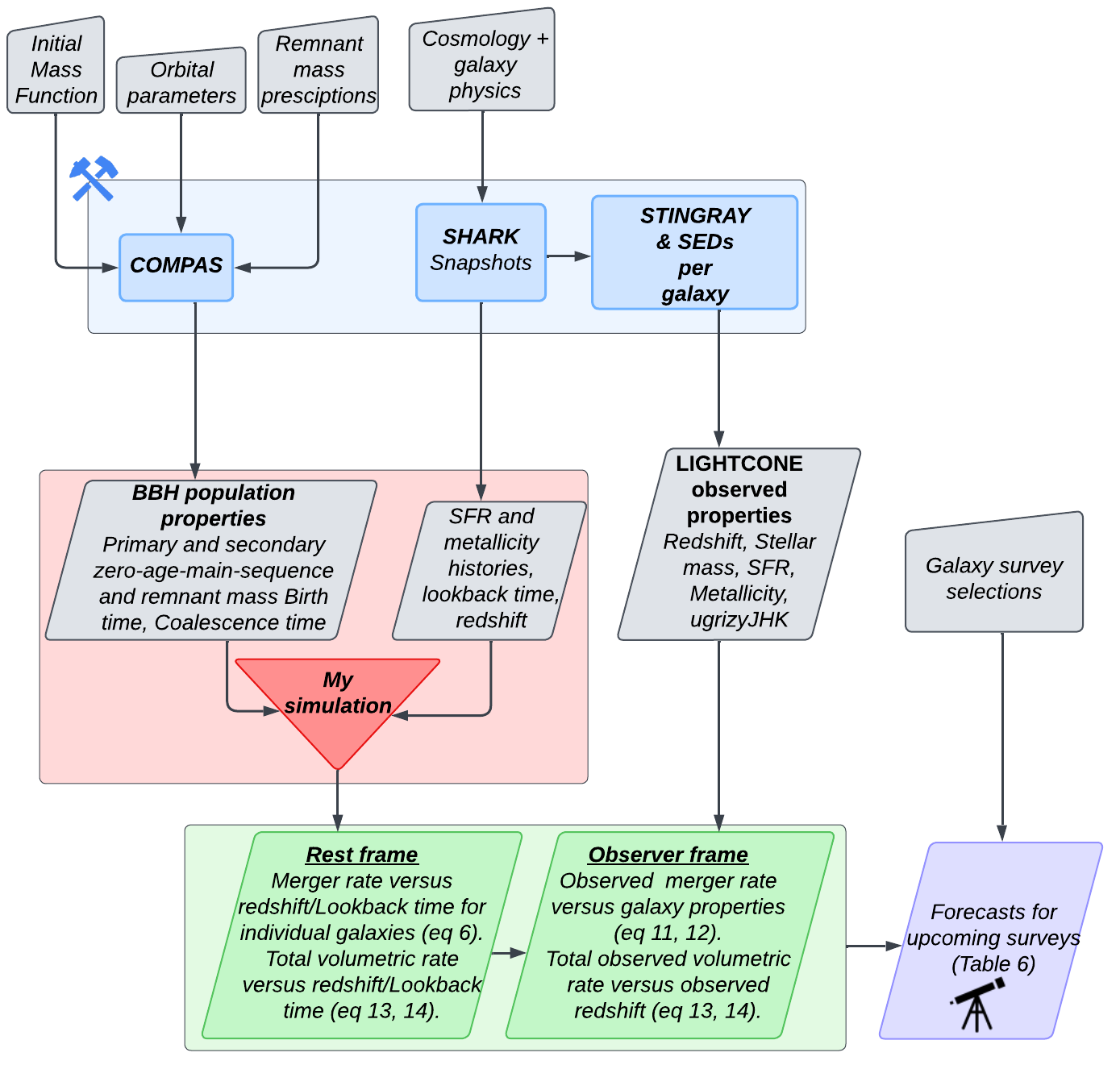}
    \caption{Flowchart outlining how the COMPAS binary population is combined with the {\sc Shark} host galaxies to determine a model for the BBH merger rate. We also show the outputs from {\sc Shark} that can be used for various applications (see Section \ref{Section 5} and \ref{Section 6}).}
    \label{fig:flowchart}
\end{figure}
\subsection{Birth and Merger rate} \label{method}
Having introduced the {\sc Shark} and COMPAS simulations, demonstrated the properties of the galaxies and binary systems they simulate, and discussed their realism and how the various choices in how they are run may change our later results, we turn to our methodology of linking these together. The main goal is to assign the COMPAS BBHs to galaxies with the appropriate metallicity in a way that ties rate of star formation in each galaxy to the binary birth rate, and hence the synthesised BBH birth and merger rate. A schematic flowchart of the steps, inputs and outputs is given in Fig.~\ref{fig:flowchart}.

We begin by generating a library of COMPAS binaries that we can use to populate the {\sc Shark} galaxies. For each of the {\sc Shark} snapshots, we randomly sample with replacement the gas metallicities of all the galaxies, and use these  as input to generate $10^6$ COMPAS binary systems using the fiducial model specifications given in Section~\ref{Section 3}. Primary masses for each binary system are drawn from a \cite{kroupa2002initial} IMF between $5$ and $200\,M_{\odot}$. Secondary masses are generated by drawing the mass ratio $q$ from the distribution of \cite{sana2012binary}. The end result is a total of 170 (time-steps) $\times$ $10^6$ binary systems with an associated metallicity that match the distribution and range of metallicities we will encounter in {\sc Shark} across the full simulation time.

\begin{figure}
    \centering
    \includegraphics[width=\columnwidth]{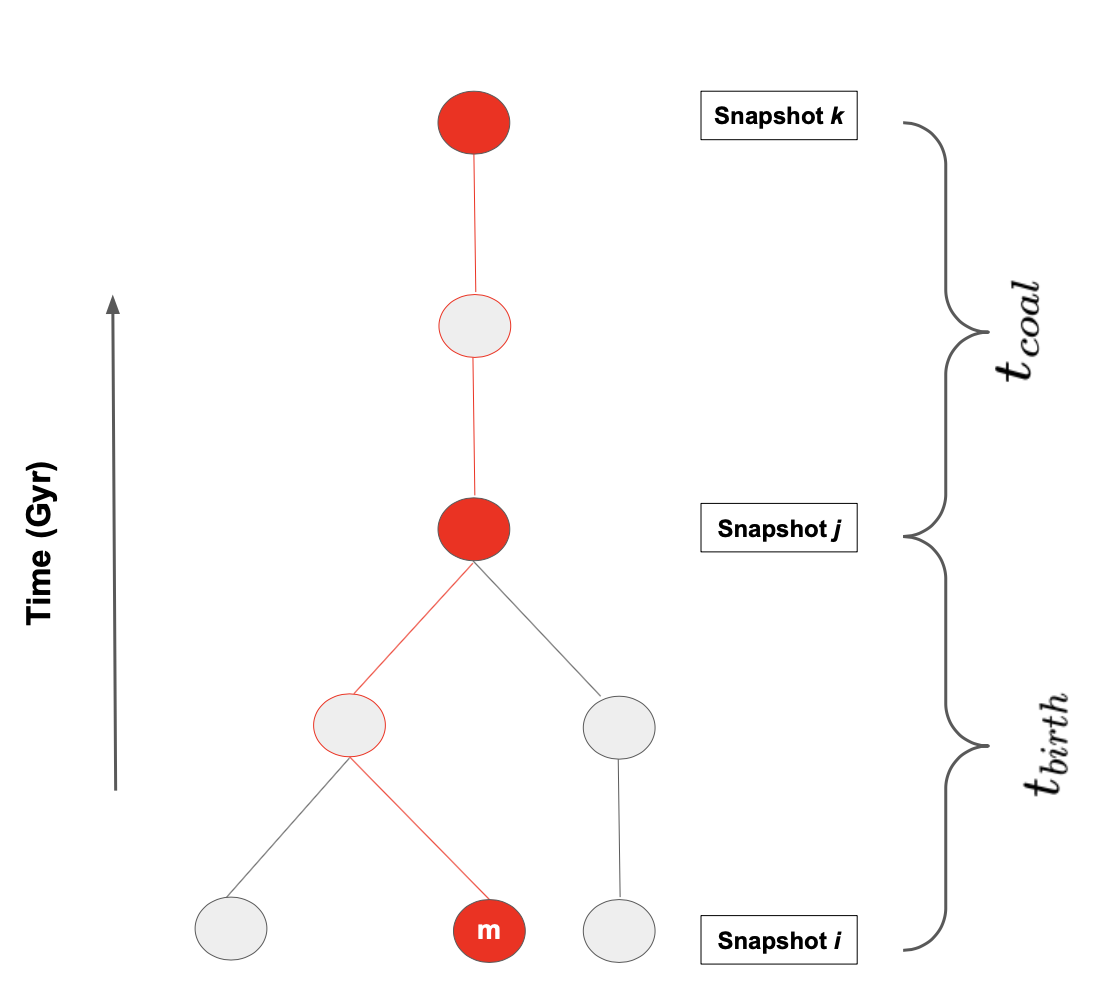}
    \caption{A simplified galaxy merger tree. Here, we follow the evolution of a single binary system along the red branch, which we denote the main branch. For this particular example, we show that a binary system formed in snapshot $i$, in its host galaxy $m$, is traced to snapshot $j$, where it evolves into a BBH system, and finally in snapshot $k$, where it merges also in galaxy $m$. This is determined using the birth and coalescence times from COMPAS.}
    \label{fig:Gal_merger}
\end{figure}

We then proceed by walking backwards down the galaxy merger tree; using the {\sc Shark} galaxy IDs at snapshot 199 ($z=0$) and the descendent galaxy IDs in snapshot 198 to find the progenitors of the galaxies in snapshot 199. We repeat this process with IDs and descendant IDs in snapshot 198 and 197 to find the progenitors for snapshot 198, and so forth. This allows us to trace the entire galaxy merger history down to the earliest snapshot in which galaxies form, and store the star formation and metallicity history for all progenitors. An example simplified merger tree is given in Fig.~\ref{fig:Gal_merger}.

Now, we apply a forward method, starting from the earliest snapshot containing galaxies (snapshot 30; $z=14.4$), and populate the progenitor galaxies with binaries from COMPAS as follows:

\begin{itemize}
    \item{For each {\sc Shark} galaxy, labelled $m$ in Fig~\ref{fig:Gal_merger}, in snapshot $i$ with time $t_{i}$, we identify the SFR $\psi_{m}(t_i)$ and metallicity $Z_{m}(t_{i})$.} 
    \item{We draw $N_{\mathrm{binary}}=50,000-500,000$ binary systems from our COMPAS library. This is done by searching for the metallicity closest to $Z_{m}(t_{i})$ and taking $N_{\mathrm{binary}}$ around this metallicity. We then return the time predicted by COMPAS for the two stars to convert into a BBH system, $t_{\mathrm{birth}}$, and for them to coalesce, $t_{\mathrm{coalescence}}$ for this $N_{\mathrm{binary}}$. Since the input parameters are fixed, COMPAS does not output a grid from which we can interpolate the BBH properties for each galaxy. However, sampling from a distribution centred around each galaxy's metallicity is more realistic.}
    \item{We now construct an array with $N_{\mathrm{binary}}$ BBH formation times given by $t_i + t_{\mathrm{birth}}$. Comparing each element in this array to the simulation times associated with each snapshot, we determine in which snapshot $j$, where $j \geq i$, the binary systems evolve into BBHs. Note that the galaxy $m$ itself may no longer exist at this point, having merged to form a different galaxy. Nonetheless, we are always able to use our merger tree to trace the ID of the galaxy in which each BBH system now resides. We store the snapshot in which each of the $50,000-500,000$ binary systems become BBH systems as a second array. Binary systems which COMPAS says will not evolve into BBHs (i.e., because one or both of their masses is too low, metallicities too high, or the binary system is disrupted at some point in its evolution), are given values of -1 and excluded from all subsequent calculations using this array.}
    \item{We apply this same method to determine when the BBHs merge using $t_{\mathrm{coalescence}}$ from COMPAS, labelled as snapshot $k$ in Fig.~\ref{fig:Gal_merger}, where $k \geq j$,  In this case we determine if $t_i + t_{\mathrm{birth}} + t_{\mathrm{coalescence}}$ falls within the range ($t_k - \frac{\Delta t_k}{2}$, $t_k + \frac{\Delta t_k}{2}$) for the merger to occur in snapshot $k$. We again store all $N_{\mathrm{binary}}$ values of $k$ for each galaxy. BBH mergers which occur after the simulation ends such that, $t_i + t_{\mathrm{birth}} + t_{\mathrm{coalescence}} > t_{\mathrm{age}}(z=0)$ are given $k=-1$ such that they are removed from subsequent calculations.}
    \item{We repeat this process by populating all galaxies in snapshot $i$ before repeating for snapshot $i+1$ and so on.}
    \item{To find the total contribution to galaxy $m$ in snapshot $k$, we find the merger rate over time for the other branches that merge onto the main branch and take the sum. The output from the simulation described by Fig.~\ref{fig:flowchart} gives us the BBH mass formed per year. To determine what fraction of this BBH mass contributes to GW events and obtain a number per year, we normalise the BBH formation rate by the total mass in stars formed at the time the binary stars formed.}
\end{itemize}

Note that galaxies exist earlier than snapshot 30, however we find that the merger rate earlier than $z=10$ is negligible due to the low SFR (see Section~\ref{Section 5} for further explanation). 
Due to the varying coalescence times for different systems, the total BBH formation and merger rates in galaxy $m$ at times $t_{j}$ and $t_{k}$, which we denote $R_{\mathrm{BBH},j,m}$ and $R_{\mathrm{GW},k,m}$ respectively, requires knowledge of all prior snapshots --- two pairs of BBHs forming in the same galaxy in snapshot $j$ may have originated from stars formed at different times, or even in different progenitor galaxies. 

For the purposes of computing the birth and merger rates we take our arrays containing the BBH formation and merger times and produce histograms of the number of entries which fall within each snapshot. In order to account for the rate at which a given galaxy is actually forming stars, this histogram is weighted by the SFR, $\psi_{m}(t_{i})$. We also normalise by the total amount of stellar mass we are generating in our procedure $M_{\mathrm{tot}}$, but in such a way as to remove the dependence on the number of binary stars we assigned from COMPAS, $N_{\mathrm{binary}}$ (see Section~\ref{Norm_factor}). Mathematically, we can denote this procedure, and the number of BBHs being formed, as

\begin{equation}
     N_{\mathrm{BBH}, j, m} = \sum_{i=0}^j \frac{\psi_{m}(t_{i})}{M_{\rm tot}}\Delta t_{i} \sum_{n=0}^{N_{\rm binary}} \vartheta (t_{i}+t_{\mathrm{birth},n}, t_{j}, \Delta t_{j})~, \label{eq:NBBH}
\end{equation}
where, strictly, $t_{\mathrm{birth},n}$ is a function of $Z_{m}(t_{i})$ (as is $t_{\mathrm{coalescence}}$), but we have dropped the dependence on metallicity here for succinctness, and 
\begin{align}
\Delta t_j &= t_j - t_{j-1} \\ 
\vartheta (\tau, t_j, \Delta t_j) &= \left\{ \begin{array}{cc} 
                 1 & \hspace{1.5mm} \text{if $t_j - \frac{\Delta t_j}{2} < \tau < t_j + \frac{\Delta t_j}{2}$,} 
\\
                0 & \hspace{1.5mm} \text{otherwise.} \\
\end{array} \right\} \label{eq:weight_2}
\end{align}
Similarly, the number of mergers can be taken by convolving the birth rate above with the coalescence time,
\begin{equation} 
    N_{\mathrm{GW}, k, m} = \sum_{i=0}^k \frac{\psi_{m}(t_{i})}{M_{\rm tot}}\Delta t_{i} \sum_{n=0}^{N_{\rm binary}} \vartheta (t_{i}+t_{\mathrm{birth},n}+t_{\mathrm{coalescence},n}, t_{k}, \Delta t_{k})~, \label{eq:NGW}
\end{equation}
To convert from the number of BBHs or GWs forming in a given snapshot to a rate, we simply divide by the width of the snapshot, i.e., 
\begin{equation}
R_{\mathrm{BBH}, j, m} = N_{\mathrm{BBH}, j, m}/\Delta t_{j}. \qquad R_{\mathrm{GW}, k, m} = N_{\mathrm{GW}, k, m}/\Delta t_{k}. \label{eq:new_model}
\end{equation}
Producing the rates this way, computing first the number of BBHs or mergers by scaling out the dependence on the width of snapshot $i$ in Eqs.~\ref{eq:NBBH} and \ref{eq:NGW} and then dividing later by the timestep for snapshots $j$ and $k$ ensures we account properly for the fact that the snapshots in {\sc Shark} are not uniform widths in lookback time.

\subsection{Normalisation factor} \label{Norm_factor}
Our calculation for the number of BBHs and GWs being formed in a given galaxy requires us to divide the SFR by the total amount of stellar mass being created, in such a way that we cancel out the dependence on the number of stars we are actually drawing from our COMPAS runs ($N_{\mathrm{binary}}$). In a sense, we need to determine how a unit star-forming mass would be distributed into individual binary systems to ensure our rates are correctly normalised.

COMPAS already partially accounts for this, in that it generates binary pairs with masses drawn from an initial mass function and mass ratio distribution specified when the code is run (our fiducial simulation uses the \cite{kroupa2002initial} IMF\footnote{{\sc SHARK} uses the \cite{chabrier2003galactic} IMF to determine the chemical enrichment in the ISM. This may imply there is a mismatch between the COMPAS binaries and the metallicity distribution in the galaxies, thus preventing us from populating the {\sc Shark} galaxies with COMPAS binaries. However, the \cite{kroupa2002initial} and \cite{chabrier2003galactic} IMF are the same for $M_{\star}> 1M_\odot$ so no calibration is required.} and \citealt{sana2012binary} ratio distribution), and masses are drawn from the IMF between a minimum and maximum mass of $5\,M_{\odot}$ and $200\,M_{\odot}$ respectively. However, in the real Universe, stars are also born singly, and with masses outside the above limits (particularly with masses below $5\,M_{\odot}$). We do not include such systems in our COMPAS runs, as they will not contribute to the BBH GW rate, and so do not need to be evolved. However, the fact that a large fraction of the stars that would be formed by each galaxy in our simulation are not represented by our COMPAS runs needs to be included as a normalisation factor, which we call $M_{\mathrm{tot}}$, in order for the overall rate of BBH formation and mergers to be realistic.

We start by writing down the analytic solution to the total mass of stars a galaxy could form, including single systems, as a function of the stellar mass ranging from 0 to infinity, binary mass ratio distribution ($\phi(q)$), Kroupa IMF ($\phi(M)$) and fraction of stars that form in binary systems ($f_{\mathrm{bin}}$). In integral form, this can be expressed as
\begin{align} \label{eq:1}
    M_{\rm ideal} &= \frac{f_{\rm bin}}{2}\int_0^{\infty} \int_0^1 (M + q \times M)\phi(M)\phi(q) dM dq \nonumber \\
    &+ (1-f_{\rm bin}) \int_0^{\infty} M \phi(M) dM~.
\end{align}
If we assume a power law for the mass ratio distribution, $\phi(q) = \beta q^{\beta-1}$, as in \cite{sana2012binary}, we can simplify this to
\begin{equation}
    M_{\rm ideal} = \int_0^{\infty} M \phi(M) \delta M \left(1 - \frac{f_{\rm bin}}{2} \frac{1}{1 + \beta}\right). 
    \label{eq:massideal}
\end{equation}

Second, we consider the total amount of mass we are assigning to a galaxy by cross-matching to our COMPAS runs, which ranges from $5M_{\odot}$ to $200M_{\odot}$. We can write the total mass from the COMPAS run, which only produces binaries ($f_{\rm bin} = 1$), as, \par 
\begin{align} \label{eq:2}
    M_{\rm COMPAS} &= \sum_i^{N_{\rm binary}} M_{i, 1} + M_{i, 2} \nonumber \\
    &= \frac{1}{2}\int_{5M_\odot}^{200M_\odot} M \phi(M) \delta M \left( [q^{\beta}]_{0.01}^1 + \frac{\beta}{1 + \beta} [q^{1 + \beta}]_{0.01}^1 \right)
\end{align}
where the first equation arises from our actual sampling of the COMPAS binaries, whose individual masses drawn from the IMF are denoted $M_{i}$, and the second equation arises from our theoretical expectations for this sampling given Eq.~\ref{eq:massideal} and the mass and mass-ratio limits of our COMPAS runs. Finally, to remove the dependence on $N_{\rm binary}$ from our rates, we can define the total stellar mass generated by a galaxy $M_{\mathrm{tot}}$ not as $M_{\mathrm{ideal}}$, but rather as the multiplication of $M_{\mathrm{ideal}}$ with the ratio of the two forms for $M_{\mathrm{COMPAS}}$ given above. This gives the following,
\begin{align}
    M_{\rm tot} = \frac{\sum_i^{N_{\rm binary}} (M_{i, 1} + M_{i, 2}) \int_0^{\infty} M \phi(M) \delta M \left(2 - \frac{1}{1 + \beta} f_{\rm bin}\right)}{\int_{5M_\odot}^{200M_\odot} M \phi(M) \delta M \left( [q^{\beta}]_{0.01}^1 + \frac{\beta}{1 + \beta} [q^{1 + \beta}]_{0.01}^1 \right)}.
\end{align}
Now, increasing $N_{\rm binary}$ increases the number of BBHs formed in each snapshot, but it also increases the normalisation factor in a consistent manner since it depends on the sum of the COMPAS masses. Given we adopt the \cite{sana2012binary} mass ratio distribution, which sets $\beta=-2.35$, there is only one free parameter. This is $f_{\rm bin}$, the fraction of stars in binaries, which was also identified in Section~\ref{sec:feff}. For the purposes of the results in this paper we  set $f_{\rm bin}=0.7$ \citep{neijssel2019effect}.

\begin{figure*}
    \centering
    \includegraphics[width=\textwidth]{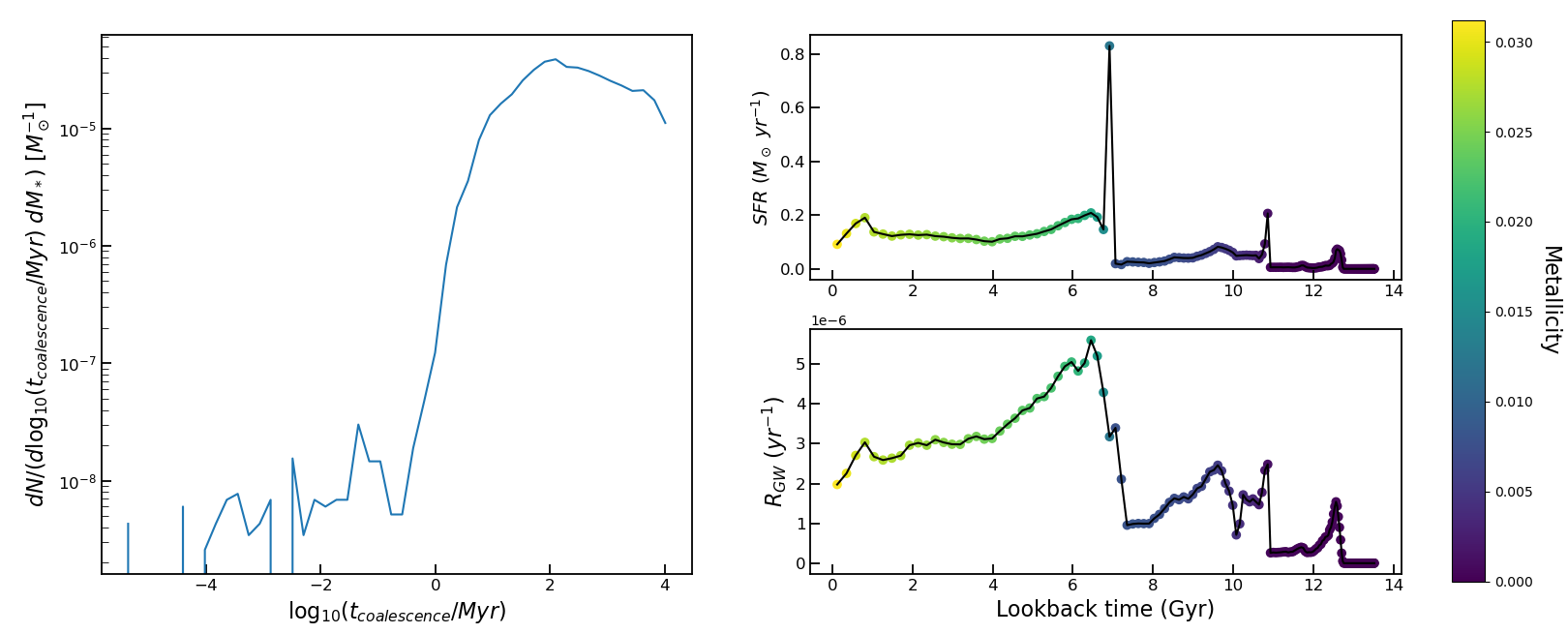}
    \caption{A visual summary of our computational GW model for a single galaxy. Taking the SFR (top right) and metallicity history from {\sc Shark} and convolving with the COMPAS coalescence times for that metallicity (right) gives us the BBH merger rate (bottom right). The SFR and merger rate are coloured by the metallicity in each snapshot, which is given by the colour bar, and the coalescence time plot shows the frequency of coalescence times per star forming mass and time-step. The SFR history shows the combined SFR from the disk and bulge, although we are also able to track which component the stars are actually born and merge in. We have combined all the coalescence times from the binaries born in this galaxy throughout the simulation to make the distribution in the left plot.}
    \label{fig:single_gal}
\end{figure*}

\subsection{Summary and lightcones} \label{lightcone}
Fig.~\ref{fig:single_gal} summarises the inputs/output for a single example galaxy after implementing the method outlined in this section. The SFR and coalescence times are effectively convolved to give the GW rate, denoted $R_{\rm GW}$, as a function of lookback time. As a result of the convolution, the merger rate maintains the overall shape of the SFR, but the sharp features are smoothed and stretched out (note for example the peaks at around 6~Gyr and 11~Gyr, which are bursts of star formation, and result in similar increases in the GW rate). The sharp increases in GW rate are almost coincident with the increases in SFR rate due to the short delay time for the majority of mergers (the peak coalescence time is around 100~Myr), but the longer tail of delay times cause the GW rate to decline far less rapidly than the SFR following the burst. This same effect causes the merger rate in general to be shifted to the left (closer to redshift 0) compared to the SFR.

Taking the tabulated merger rate as a function of lookback time for each galaxy in our snapshot, we can determine the merger rate of the observed galaxies in a lightcone. The {\sc Shark} lightcone also contains the simulated photometry of the galaxy and so this cross-matching will allow us to explore the relationship between various survey selection functions and the GW rate of the galaxies they observe. Firstly, we find if the galaxy exists in the lightcone by cross-matching the galaxy ID to the ID in the lightcone, and if so, take the cosmological redshift and lookback time. Then, we determine which snapshots this redshift lies between, find the corresponding merger rates in the merger rate history and linearly interpolate. We can apply this method to every galaxy in the mock to determine the relationship between the observed merger rate and the host galaxy properties at the time the merger occurred. The results of this process, looking at the observed merger rate, which we denote $R_{\rm GW, obs}$\footnote{We refer to the lightcone merger rate at the observed redshift as the observed merger rate. We are assuming a perfect observation of the host galaxies and leave incorporation of GW detector selection effects for later work.}, as a function of galaxy properties in the lightcone are explored in the remainder of this paper. 
\par 
Note that due to the stellar mass cut of $M_*/h > 10^8 M_\odot$ in our lightcone arising from the simulation resolution limit, galaxies below this cut-off are not observed. This reduces the total merger rate in each redshift bin, as can be seen later in Section \ref{Section 5}. This does not have a large effect on our results going forward however, as the majority of mergers occur in higher mass galaxies, and such low mass galaxies will not be observed with high `completeness' (except perhaps at very low redshifts) by the surveys we consider. This will be explained further in Section \ref{Section 6}.

\section{Results and Discussion} \label{Section 5} 
In this section, we investigate the observed merger rate, $R_{\rm GW, obs}$, as a function of \textit{intrinsic} host galaxy properties from the lightcone, determine the best fit and intrinsic scatter in our model for the predicted merger rate and compare to other predictions from the literature. We then compare the total rate to current observations, discuss the possible causes of uncertainties and discrepancies in our model predictions and determine the best fit for the total rate. While our initial analysis is based on the results with the \citet{fryer2012compact} remnant mass model prescription, we also include outputs for \citet{mandel2020simple}, as this model closely aligns with the predictions from GWTC-3 around $z=0$, and \cite{schneider2021pre}. 

\subsection{Merger rate vs global properties}
\begin{figure*} 
    \centering
    \includegraphics[width=\textwidth, height=7.2cm]{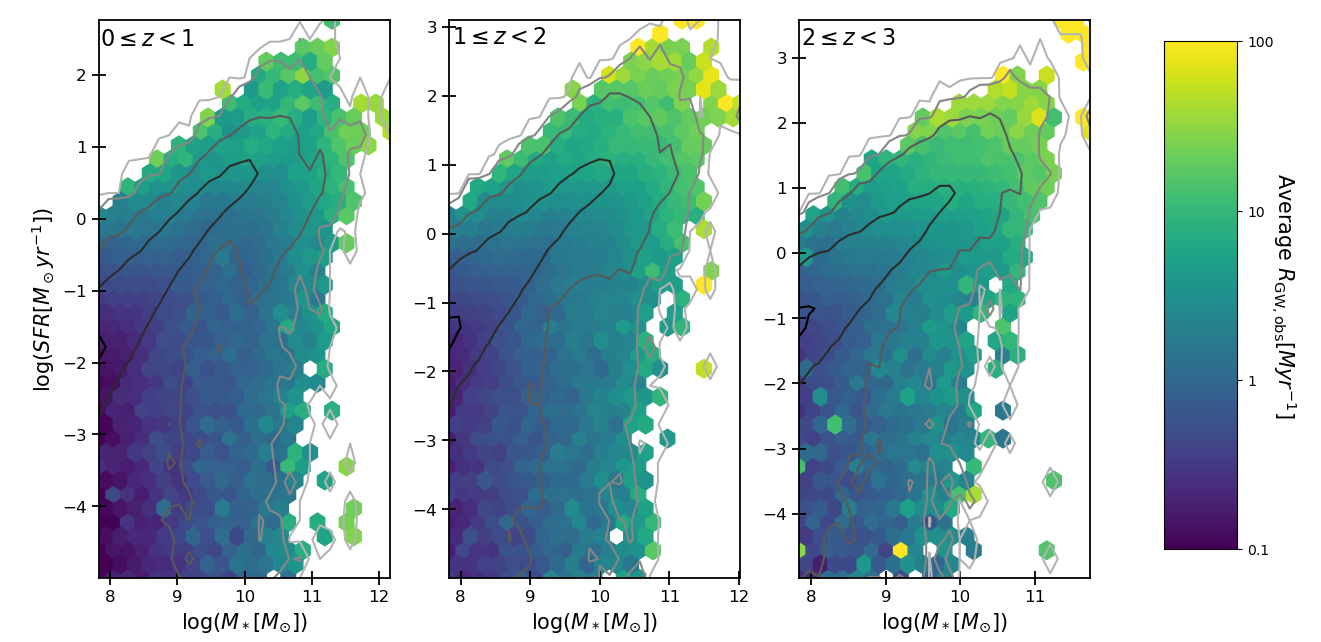}
    \includegraphics[width=\textwidth, height=7.2cm]{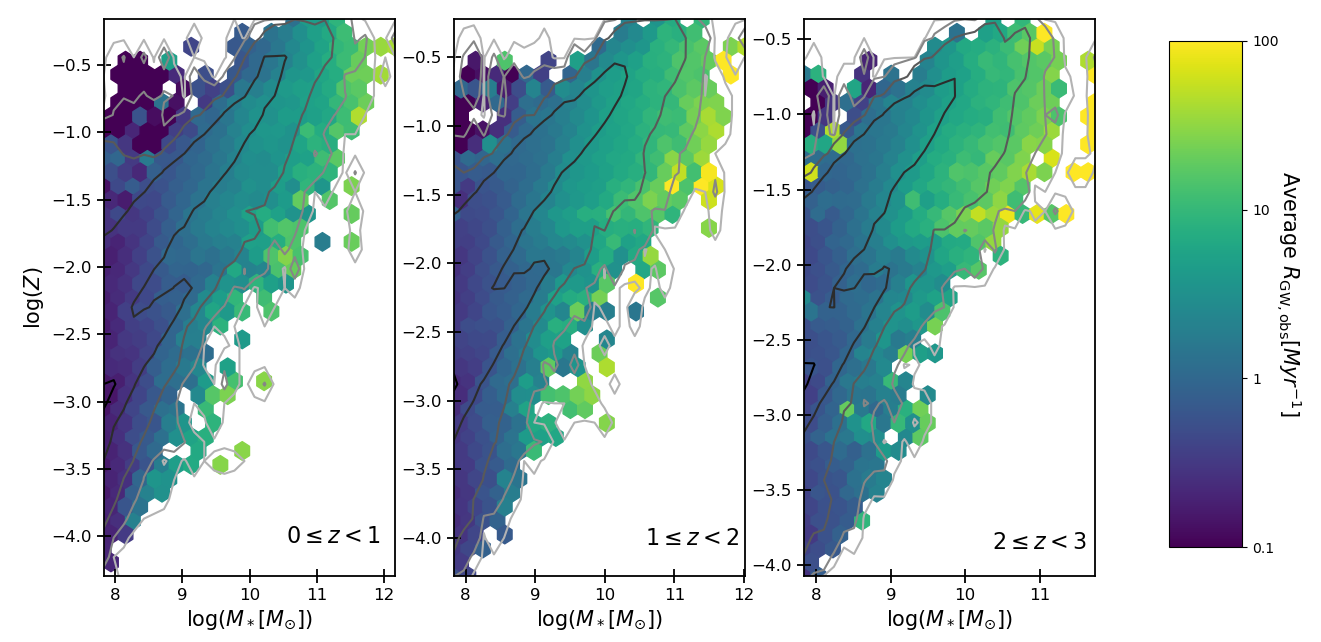}
    \includegraphics[width=\textwidth, height=7.25cm]{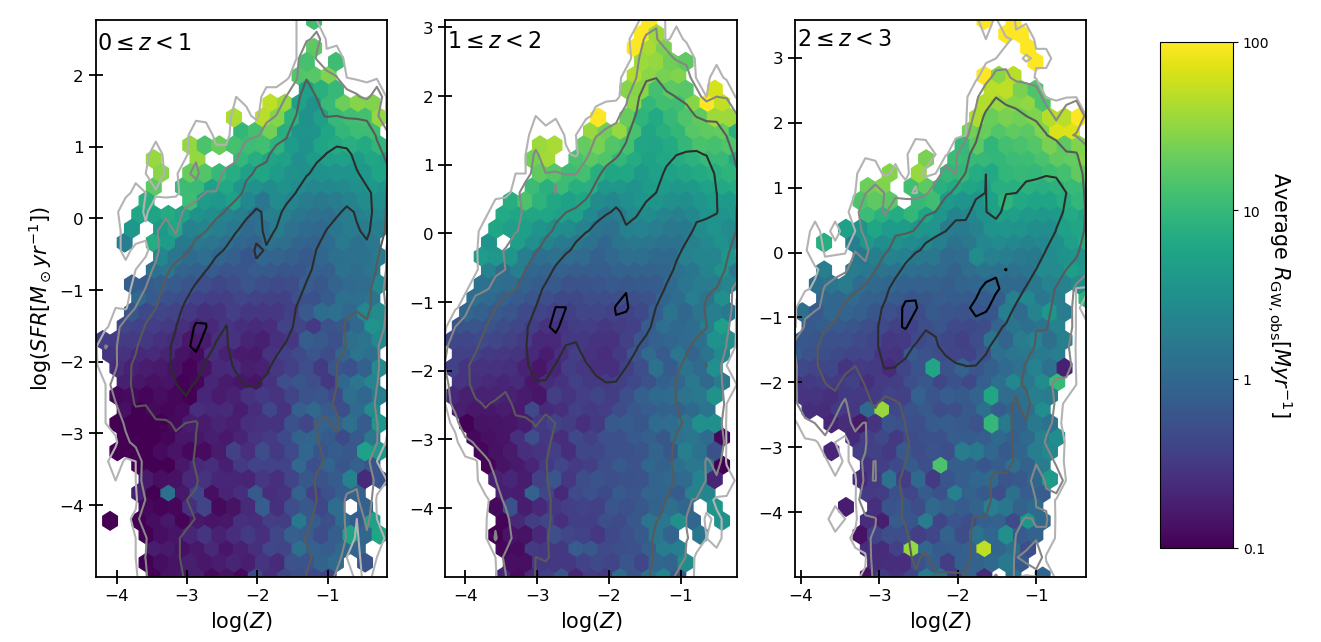}
    \caption{Plots of combinations of the fundamental galaxy properties from the lightcone; stellar mass, SFR and stellar metallicity coloured according to the average observed GW rate (refer to \ref{lightcone} for how this is found), with a $\log(\rm SFR) < -5$ cut. Contours show the number of galaxies, with the lightest coloured contour corresponding to one galaxy per bin and the darkest coloured contour ranging from 5,000 (in the plot $\log(M_*)$ vs $\log(\rm Z)$) to 35,000 (in the plot $\log(M_*)$ vs $\log(\rm SFR)$) per bin. For every row, each frame represents an observed redshift range; $0 \leq z < 1$, $1 \leq z < 2$ and $2 \leq z < 3$. For each bin, the $\log(\rm SFR) < -5$ cut removes 5.6\%, 2.2\% and 0.6\% of the galaxy sample respectively.}
    \label{fig:host_properties}
\end{figure*}
\begin{figure*}
    \centering
    \includegraphics[width=\textwidth, height=7.2cm]{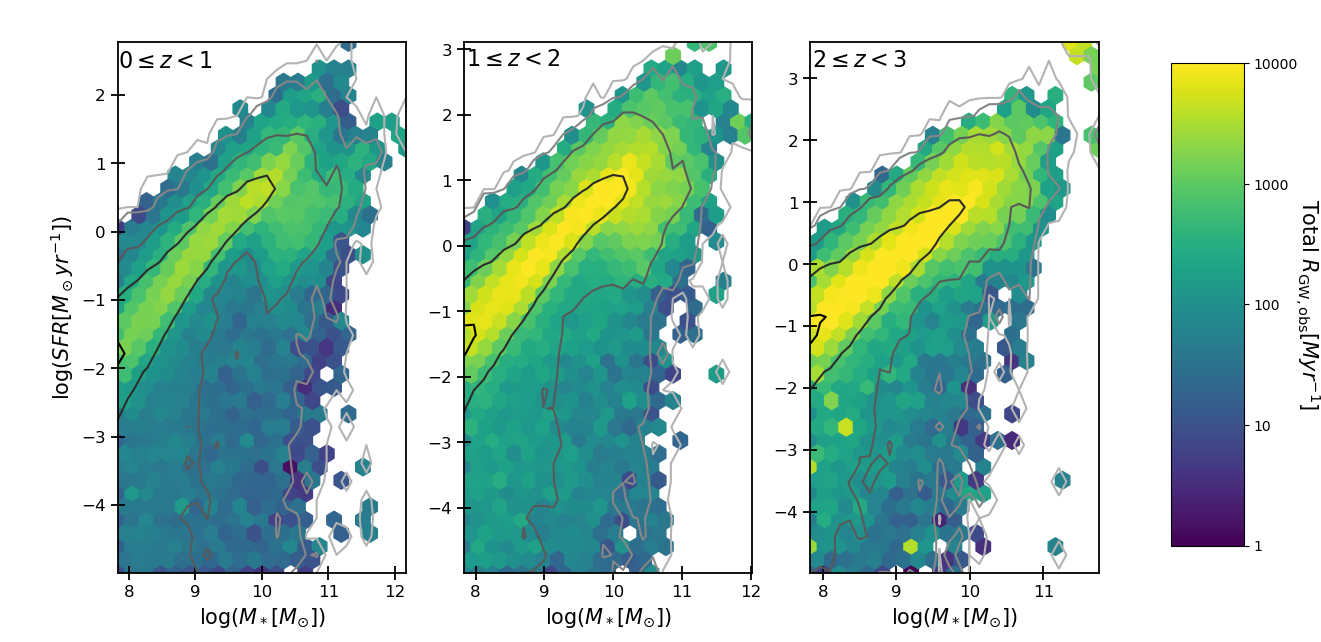}
    \includegraphics[width=\textwidth, height=7.2cm]{Muller_Mandel_plots/M_stellar_vs_SFR_SFR_cut_total.png}
    \includegraphics[width=\textwidth, height=7.25cm]{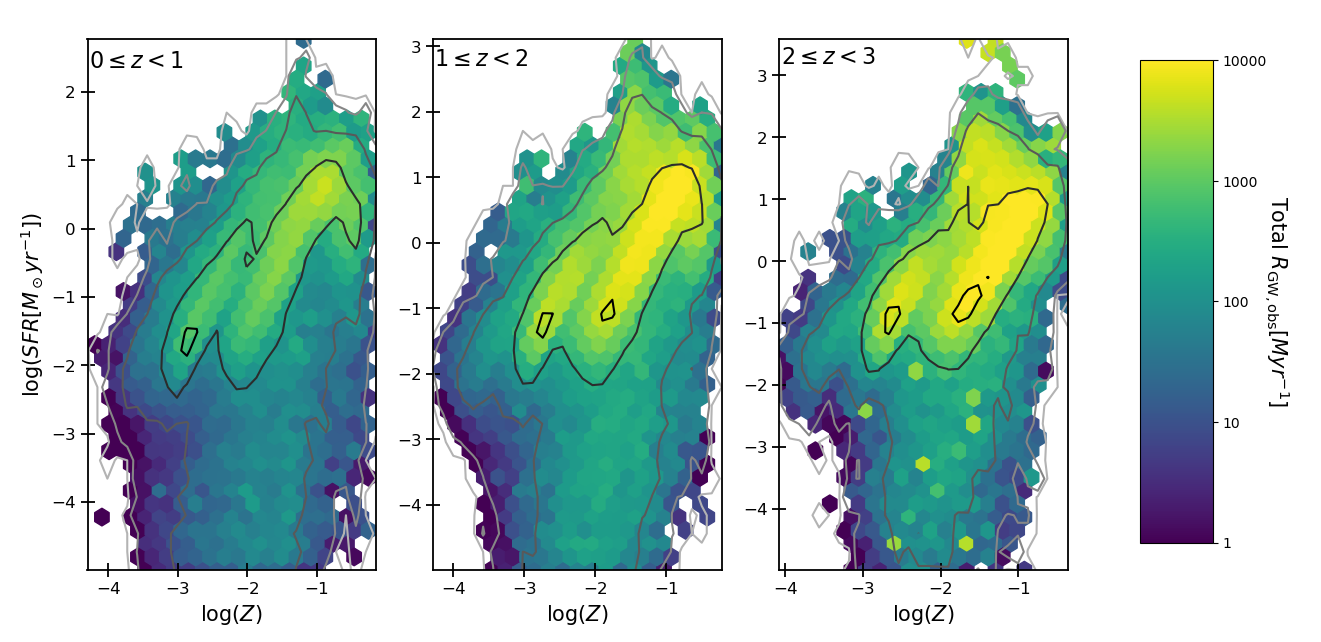}
    \caption{Same as Fig~\ref{fig:host_properties} except the hexbins are coloured by the \textit{total} merger rate.}
    \label{fig:host_properties_total}
\end{figure*}

In Fig~\ref{fig:host_properties} and Fig~\ref{fig:host_properties_total}, we bin the galaxies in the lightcone as a function of stellar mass, SFR and stellar metallicity in redshift slices, and for each bin we then compute the average and total number of galaxies respectively. Our main findings are as follows: 

\begin{itemize}
\item{There is a strong correlation between the stellar mass and GW rate. Galaxies with larger present day stellar mass must have had more star formation in the past, or continually formed stars throughout the simulation. Given that the same IMF is assumed for all galaxies, a larger number of stars and BHs would have formed in these high stellar mass galaxies. More BBHs lead to more mergers, and hence, more GW events.}



\item{For galaxies on the star forming main sequence, at fixed stellar mass we see evidence for higher GW rate in galaxies with higher observed star formation rate. This residual correlation with observed SFR hence arises because the typical coalescence times for BBH mergers is short (which can be verified from Fig.~\ref{fig:COMPAS_dist}), and, crucially, shorter than the timescales for mergers or feedback. For such a correlation to exist, gas-rich mergers that lead to bursts of star formation must lead to more stars being born, converting to black holes, \textit{and} merging \textit{before} the galaxy returns to normal star formation.}

\item{Shark central galaxies are almost always actively star-forming and on the main-sequence. The tail of galaxies with $\log(\rm SFR) < -3$ corresponds to the satellite dwarf galaxies that are affected by their environment, thus resulting in SF quenching and low observed merger rates.}

\item{Finally, for lower mass galaxies in particular, we observe a trend of increasing GW rate with decreasing metallicity at both fixed stellar mass and fixed SFR. From Fig.~\ref{fig:COMPAS_dist}, the coalescence time distribution also shows that metal poor stars produce more BHs and systems with shorter coalescence times leading to greater GW rates. Inversely, massive, metal rich galaxies are unlikely to produce GW events that merge within the simulation time, which makes evidence of any residual trend with metallicity less clear.} 


\end{itemize}

From looking at the average GW rate, we find that the most likely hosts are high stellar mass, high star-forming galaxies. However, as Fig.~\ref{fig:host_properties} is focused on the \textit{average} GW rate per galaxy, it is possible that the \textit{total} number of GW mergers is dominated by the greater population of systems with lower mass and SFR. By binning the total merger rate as shown in Fig.~\ref{fig:host_properties_total}, we find that in practice there is some balance between rate per galaxy and total numbers of galaxies at a given stellar mass or SFR, such that galaxies situated cleanly on the star-formation main sequence at a given redshift are the most prolific hosts.


To provide a useful summary of the relationship between the intrinsic host galaxy properties and the merger rate, and allow for comparison with  \citet{artale2019host}, we use the Hyperfit package \citep{2015PASA...32...33R} to obtain a linear fit in the form of, 
\begin{align} 
    \log(R_{GW}/ \rm yr^{-1}) &= a\log(\rm M_*/M_\odot) + b\log(\rm SFR/M_\odot \rm yr^{-1}) + \nonumber \\ 
    c\log(\rm Z) + R_0, 
    \label{eq:prediction}
\end{align}
where we also allow for intrinsic scatter in the relationship. We also expand this model to include the dependence on the redshift, 
\begin{align} 
    \log(R_{GW}/ \rm yr^{-1}) = a\log(\rm M_*/M_\odot) + b\log(\rm SFR/M_\odot \rm yr^{-1}) + \nonumber \\
    c\log(\rm Z) + d.\textit{z} + R_0. 
    \label{eq:prediction_2}
\end{align}

\begin{figure} 
    \centering  
    \includegraphics[width=1.05\linewidth]{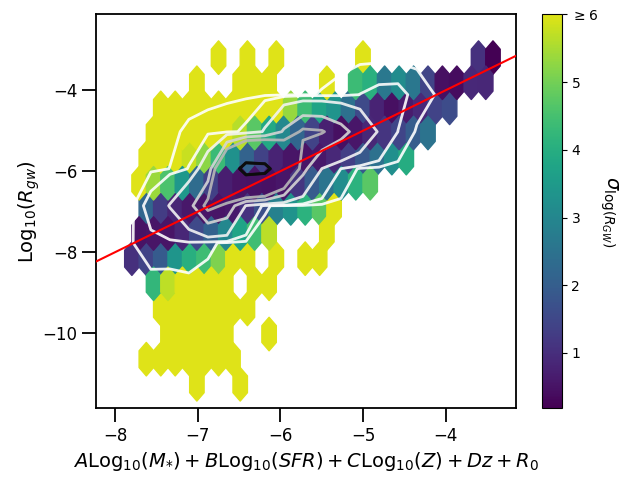}
    \caption{The predicted against the actual merger rate for $\log(\rm SFR) > -5$ and all redshifts using Eq.~\ref{eq:prediction_2} (fits given in table  \ref{tab:global_fits}). The hexbins are coloured by the average sigma value and each contour represents the distribution of 25\% (darkest), 40\%, 70\%, 87.5\%, 95\%, 98.5\% and 99.7\% (lightest) of the total galaxies in the parameter space. The red line shows the one-to-one line, where the prediction matches the actual value obtained from the model. We include a maximum cut of 6 in the colour bar, which is reasonable because galaxies with $\sigma > 6$ are outside of the highest contour range. } 
    \label{fig:fits}
\end{figure}

We perform separate fits for galaxies with $\log(\rm SFR/M_\odot \rm yr^{-1}) > -5$. The best-fitting parameters are summarised in Table~\ref{tab:global_fits}. In Fig.~\ref{fig:fits}, we plot the predicted merger rate based on the observed galaxy properties against the observed merger rate and find our model provides an excellent fit, although some residual scatter remains arising from the fact that galaxy star formation is stochastic and so galaxy properties measured at the time of the GW event do not correlate perfectly with the properties at the time the binary system forms, which ultimately determines its fate. We avoid fitting for galaxies with $\log(\rm SFR/M_\odot \rm yr^{-1}) \leq -5$. The exact SFR values below this cut are difficult to measure and cannot be trusted due to resolution limits. Thus their merger rate fits cannot be trusted naively, and we expect these fits to change with higher resolution simulations. 

\begin{table*}
    \centering
    \begin{tabular}{|c|c|c|c|c|c|c|c|}
        \hline  
        \hline
         & & a & b & c & d & $R_0$ & $\sigma_{\log(\rm RGW)}$ \\
         \hline 
        \multirow{3}{*}{$\log(\rm SFR) > -5$} & $0 \leq z < 1$ &  0.702 $\pm$ 0.002 &  0.245 $\pm$ 0.001 & -0.418 $\pm$ 0.002  & - & -13.06 $\pm$ 0.03 & 0.319 $\pm$ 0.001  \\
      & $1 \leq z < 2$ & 0.503 $\pm$ 0.001 & 0.2376 $\pm$ 0.0004 & -0.159 $\pm$ 0.001 & - & -10.62 $\pm$ 0.01 & 0.2458 $\pm$ 0.0003 \\
      & $2 \leq z < 3$ & 0.435 $\pm$ 0.001 & 0.2355 $\pm$ 0.0005 & -0.068 $\pm$ 0.001 & - & -9.70 $\pm$ 0.01 & 0.2013 $\pm$ 0.0002 \\  
        \hline  
        \hline
         $\log(\rm SFR) > -5$ & $0 \leq z < 3$ &  0.527 $\pm$ 0.001 &  0.2238 $\pm$ 0.0003 & -0.169 $\pm$ 0.001 & 0.2360 $\pm$ 0.0004 & -11.24 $\pm$ 0.01 & 0.2378 $\pm$ 0.0002 \\
        \hline
        \hline  
        \multirow{3}{*}{\citet{artale2019mass}} & $z=0.1$ & $0.921 \pm 0.001$ & -$0.051 \pm 0.001$ & -$0.404 \pm 0.003$ & - &  -$15.049 \pm 0.018$ & - \\
        & $z=1$ & $1.134 \pm 0.002$ & -$0.172 \pm 0.002$ & -$0.839 \pm 0.004$ & - & -$17.338 \pm 0.024$ & - \\
        & $z=2$ & $1.135 \pm 0.002$ & -$0.167 \pm 0.001$ & -$0.681 \pm 0.003$ & - & -$16.758 \pm 0.020$ & - \\
        \hline
    \end{tabular}
    \caption{Results of the fit to the GW rate versus galaxy properties (Eq.~\ref{eq:prediction}) and the results when redshift dependence is added (Eq.~\ref{eq:prediction_2}) for the \citet{muller2016simple} model, and the \citet{artale2019mass} fit without redshift dependence.}
    \label{tab:global_fits}
\end{table*}  

\subsection{Comparing to Observations} 
\begin{table}
	\centering
	\caption{Observed BBH merger rate density at $z=0$ ($\text{Gpc}^{-3} \text{yr}^{-1}$) from released GWTCs \citep{abbott2016binary, abbott2019gwtc, 2020LRR....23....3A, 2021arXiv211103634T} and from varying remnant mass models in COMPAS.}
	\label{tab:cases_lit}
	\begin{tabular}{ccc} 
		\hline
		\hline
		Observing run & \multicolumn{2}{c}{$R_{\rm GW}$} \\
		\hline
		LIGO O1 & \multicolumn{2}{c}{$9-240$} \\ 
		LIGO O1 + O2 & \multicolumn{2}{c}{$19.7^{+57.3}_{-15.9}$} \\
		LIGO O1 + O2 + O3a & \multicolumn{2}{c}{$23.9^{14.3}_{-8.6}$} \\
		LIGO O1 + O2 + O3a + O3b & \multicolumn{2}{c}{$18^{+43}_{-2}$} \\
		\hline
		Model & $R_{\rm GW}$ & $R_{\rm GW}$  \\
		&  (Snapshot) & (Lightcone) \\ 
		\hline 
		\citet{fryer2012compact} & $123.0_{-4.1}^{+2.6}$ & $103.0_{-8.4}^{+5.5}$\\
		\citet{mandel2020simple} & $17.7_{-0.6}^{+0.4}$ & $14.3_{-1.2}^{+0.8}$  \\
		\citet{mandel2020simple} ($f_{\rm WR} =0.2$) & $20.8_{-0.7}^{+0.4}$ & $17.4_{-1.4}^{+0.9}$ \\ 
		\citet{schneider2021pre} & $1.7_{-0.00003}^{+0.00007}$ & $1.4_{-0.1}^{+0.1}$ \\ 
		\hline
	\end{tabular}
\end{table}

\begin{figure*}
    \begin{multicols}{2}
    \centering
    \includegraphics[width=\linewidth]{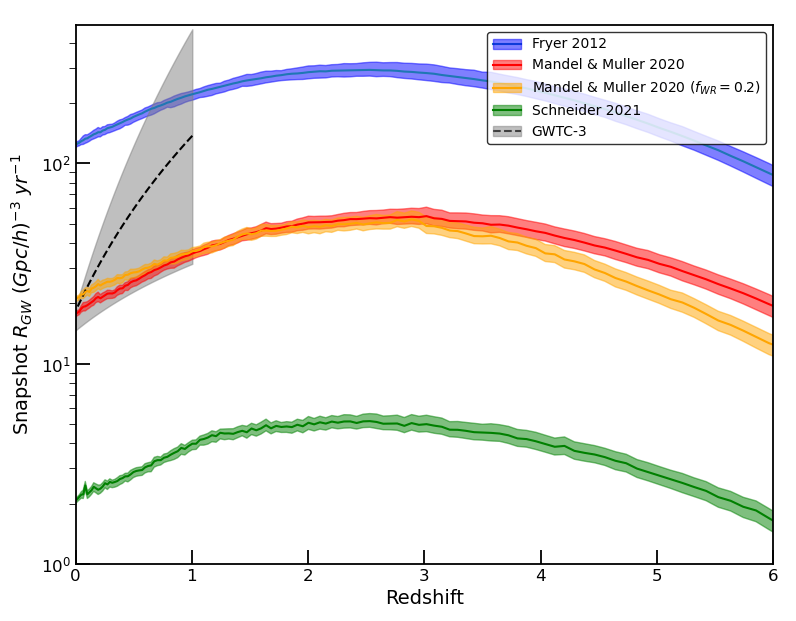} \par
    \includegraphics[width=\linewidth]{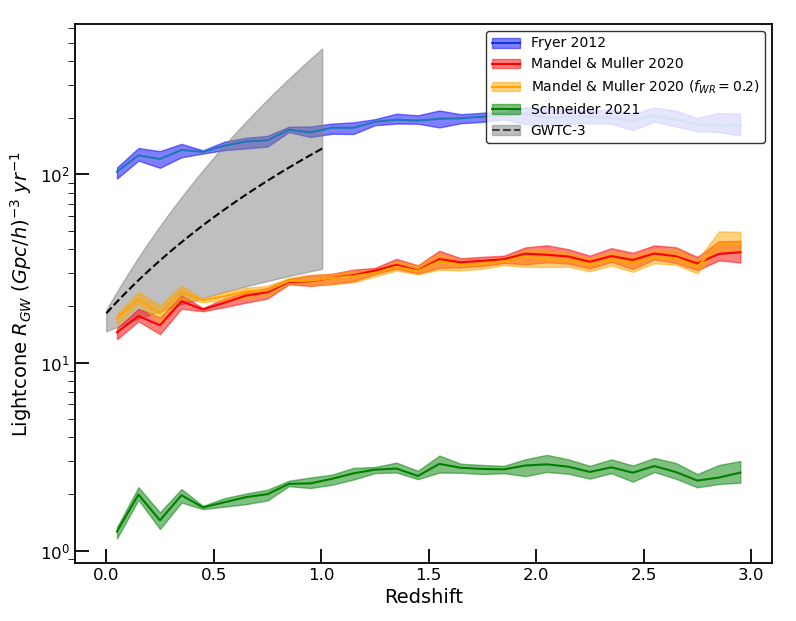}
    \end{multicols}
    \caption{Evolution of BBH merger rate density with redshift in the snapshots (\textit{left}) and the STINGRAY lightcone (\textit{right}). The merger rate follows the evolution of the cosmic SFR density shape, peaking around $z=2.5$. The dashed line is the prediction from GTWC-3 \citep{2021arXiv211103634T} valid for $z \leq 1$. The grey shaded regions cover the uncertainty on the observational constraints. This is done by taking the range in the merger rate over five subvolumes. It is clear that there is a dramatic difference in GW rate predictions depending on the remnant mass model used.}
    \label{fig:RGWvol}
\end{figure*}
In Fig.~\ref{fig:RGWvol} we show the volumetric rate for GWs from BBH mergers using both our snapshots and the lightcone. We also show the observational fit given as a power law (based on the work by \citealt{2018ApJ...863L..41F}), where the parameter fits are determined using the most recent data from GWTC-3 \citep{ligo2021population}. Table~\ref{tab:cases_lit} summarises the observed and predicted merger rates at $z=0$. To give an indication of the error due to cosmic variance we plot our measurements as a band with range taken using 5 of our simulation subvolumes. The volumetric rate in the lightcone is slightly lower than the rate in the snapshots due to the exclusion of galaxies with $M_{\star}/h >10^8 M_\odot$ total stellar mass. 

There are two main features seen in these plots. Firstly,  our default volumetric rate with the \citet{fryer2012compact} remnant-mass model over-predicts the merger rate compared to the GWTC-3 model, particularly at low redshift, however it does fall into the range of the merger rate predicted at $z>0.5$. Secondly, the shape of our predictions, rises less steeply towards higher redshift than the extrapolation made from the GWTC-3 data. 
There are a number of ways in which the simulation could be brought into agreement with the observations. Given the majority of BBH mergers have short coalescence times, the GW rate is roughly proportional to the SFR density. This was shown previously for {\sc Shark} in Fig.~\ref{fig:CSFR}, where we see a discrepancy of $\sim0.3\mathrm{dex}$ at $z=0$ compared to the measurements from \cite{driver2018gama}. Adjusting the inputs to {\sc Shark} so that they more closely align with this data and the curve from \cite{madau2014cosmic} would potentially bring the shape of our predicted volumetric gravitational wave rate in closer agreement to the GTWC-3 constraints.   


In Section~\ref{sec:feff}, we have also investigated how different modelling choices can change the COMPAS fraction. 
If we used the more up-to-date remnant mass models from Case 7 \citep{mandel2020simple} and 8 \citep{schneider2021pre} from Table \ref{tab:cases_lit}, our rates would be more in line with the most recent predictions. \citet{mandel2020simple} (Case 7) and \cite{fryer2012compact} (Fiducial) have a similar $M_{\rm core}$ vs $M_{\rm remnant}$ relationship. However, \citet{mandel2020simple} implements a probabilistic recipe to account for stochasticity in stellar evolution, which reduces our predicted GW rate at $z=0$ by a factor of 5. \citet{schneider2021pre} (Case 8) consider common envelope stripping in their prescriptions, which leads to a smaller initial mass range for BBHs, and a reduction of the BH population by 90\%. We further explore the \citet{mandel2020simple} case by altering the Wolf-Rayet multiplier $f_{\rm WR}$, which determines the stellar wind mass loss, from the default $f_{\rm WR}=1$ to $f_{\rm WR}=0.2$. We also change the natal kick magnitude distribution, which is the push experienced by black holes born after a supernovae explosion, from the default Maxwellian \citep{hobbs2005statistical} to \citet{mandel2020simple}. Although, we expect these natal kicks to have no impact on large remnant masses. We find an increase in the merger rate density in the $0 \leq z \leq 1$ region, which fits better with the GWTC-3 prediction. This is due to the binary experiencing weaker winds, which increases the remnant mass and reduces the binary widening, leading to shorter merger times. Varying the stellar evolution model parameters is an unpredictable process and limiting ourselves to these four cases will not necessarily align the theoretical predictions with observations, but it can clearly resolve the discrepancy between our fiducial predicted rate and the GTWC-3 observations.

It is worth highlighting that a similar broad spread (over many orders of magnitude) in the volumetric rate is seen in \cite{giacobbo2018progenitors, tang2020dependence} and \cite{mandel2021rates}. In this latter work in particular, the predictions for BBH mergers via isolated binary evolution ranges between $0.4-6000$Gpc$^{-3}$yr$^{-1}$, an exceptionally wide range that also encompasses our results. Summarising these and our works, the main sources of uncertainty are in the input parameters (initial mass function, binary separation and eccentricity), and in the general metallicity to remnant-mass relationship which includes mechanisms that are not well understood such as mass transfer and the common-envelope phase. 

Our simulation shows that the specific combination of the \cite{fryer2012compact} or \cite{schneider2021pre} model with the default COMPAS input distributions to generate BBHs and GW mergers is incompatible with current data. One benefit of our method to combine semi-analytic galaxy simulations and a fast population synthesis code is that both of these can be run for different initial choices relatively quickly. This in turn could allow for better constraints on galaxy and binary evolution, particularly as uncertainty on the redshift dependence of the observed volumetric rate decreases with more GW detections. We leave a more rigorous analysis of how the GWTC-3 data constrains these via a methodology such as the one we present here for future work.

\subsection{Volumetric rate fits}
As with our predictions for the GW rate as a function of intrinsic galaxy properties and for the purposes of future comparisons/forecasting, we provide here simple fits to the volumetric rates predicted from our simulations. This may be particularly helpful above $z=1$, where the inference from the GWTC-3 data becomes limited by the lack of detections. We investigated several functional forms for the fits, but report here two, one that is expected based on the merger rate being correlated with the SFR, i.e., given in the same form as the  \citet{madau2014cosmic} SFR density,

\begin{align}
    R_{\rm GW} = R_0 \frac{(1+z)^\alpha}{1 + \left(\frac{1+z}{z_p}\right)^{\beta}} \hspace{1ex} \rm (Gpc/h)^{-3} yr^{-1},
\end{align}
with the following priors to ensure a turnover at $z_{\rm max}$, 
\begin{align}
    & \alpha \leq \beta, \nonumber \\
    & 1 + z_{\rm max} - 0.5 < z_p\left(-\frac{\alpha}{\alpha - \beta}\right)^{\frac{1}{\beta}} < 1 + z_{\rm max} + 0.5, \nonumber \\ 
    & R_{\rm GW}(z=0) - 5 < \frac{R_0}{1+\left(\frac{1}{z_p}\right)^\beta} < R_{\rm GW}(z=0) + 5, \nonumber
\end{align}
and a second fit, a skew normal distribution, which worked particularly well in capturing the turnover at higher redshifts without any conditions,

\begin{align}
    R_{\rm GW} = \frac{R_0}{\alpha}\exp\left(-\frac{1}{2}\left(\frac{z-\beta}{\alpha}\right)^2\right)\left[\erf\left(\frac{\kappa(z-\beta)}{\sqrt{2}\alpha}\right) + 1\right].
\end{align}

\begin{table*}
    \centering
    \begin{tabular}{cccccc|ccccc}
    \hline
    & \multicolumn{4}{c|}{Snapshot} & \multicolumn{4}{|c}{Lightcone} \\
    \hline \hline 
    Skew Normal & & & & & & & & \\
    \hline
    & $R_0$ & $\mu$ & $\sigma$ & $\alpha$ & $\chi^2$ & $R_0$ & $\mu$ & $\sigma$ & $\alpha$ & $\chi^2$ \\ 
    \hline
    \citet{fryer2012compact} & 607.992 & 2.906 & 0.890 & 1.622 & 0.630 & 368.714 & 2.193 & 0.914 & 0.988 & 3.902 \\ 
    \citet{mandel2020simple} & 111.878 & 2.986 & 1.117 & 1.777 & 0.533 & 123.143 & 6.126 & 0.438 & 5.626 & 1.944 \\ 
    \citet{mandel2020simple} ($f_{\rm WR} =0.2$) & 97.366 & 2.382 & 1.275 & 1.021 & 2.332 & 70.917 & 2.001 & 2.376 & 0.001 & 1.668 \\ 
    \citet{schneider2021pre} & 10.710 & 2.116 & 2.226 & 0.256 & 1.182 & 5.011 & 2.408 & 0.682 & 1.418 & 0.343 \\ 
    \hline
    \hline
    \citet{madau2014cosmic} & & & & \\
    \hline
    & $R_0$ & $\alpha$ & $\beta$ & $z_p$ & $\chi^2$ & $R_0$ & $\alpha$ & $\beta$ & $z_p$ & $\chi^2$\\ 
    \hline
    \citet{fryer2012compact} & 122.631 & 0.855 & 4.838 & 4.736 & 5.505 & 103.411 & 0.797 & 4.294 & 4.303 & 3.739 \\ 
    \citet{mandel2020simple} & 17.336 & 1.048 & 4.907 & 4.758 & 1.289 & 13.850 & 1.055 & 4.280 & 4.320 & 1.483 \\ 
    \citet{mandel2020simple} ($f_{\rm WR} =0.2$) & 20.584 & 0.842 & 4.856 & 5.464 & 1.438 & 17.181 & 0.716 & 4.738 & 5.459 & 1.704 \\ 
    \citet{schneider2021pre} & 2.029 & 0.967 & 4.592 & 4.274 & 0.452 & 1.365 & 0.839 & 4.152 & 4.651 & 0.316 \\ 
    \hline
    \end{tabular}
     \caption{Fits for the total volumetric merger rate as a function of redshift in the snapshot and lightcone for different remnant-mass models and using two different functional forms.}
     \label{tab:vol_fits}
\end{table*}

In Table \ref{tab:vol_fits}, we show the best-fitting parameters for these fitting formulae, applied to the results from different binary models. We find that the \citet{madau2014cosmic} and the right- skewed normal distribution both capture the merger rate density distribution up to redshift 6. Therefore, we recommend both fits to make a quick estimate of the GW volumetric rate versus redshift.

\section{Optical surveys: Impact of selection effects} \label{Section 6}
In the previous section we looked at predictions for the GW rate as a function of intrinsic galaxy properties. In this section we turn now to what is actually measured --- photometry. As such, we investigate the BBH merger rate `completeness', which is the ratio of the merger rate for galaxies observed in different surveys and the merger rate in the full {\sc Shark} lightcone, as a function of distance. The full lightcone is deeper than any of the surveys we consider. 

\begin{figure}
    \centering
    \includegraphics[width=1.05\columnwidth]{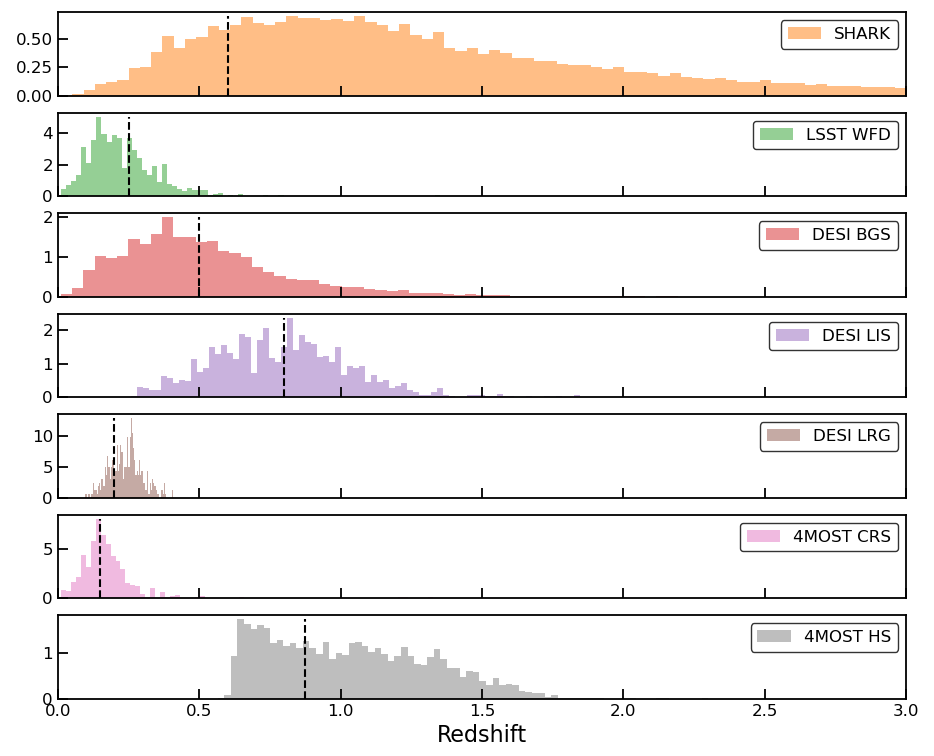}
    \caption{Normalised distribution of the {\sc Shark} observed redshifts after applying the selections for each optical survey. The dashed line shows the peak in the predicted redshift distribution for each survey from their respective survey definition references.}
    \label{fig:z_survey_dist}
\end{figure}

We anticipate this information to be useful for designing optical surveys that will target galaxies with high merger rates, as well as understanding which upcoming surveys are most useful for GW host identification. Note that the GW events will be detected regardless of a host galaxy being found or not. However, GW cosmological measurements are limited by galaxy catalogues covering only a fraction of the sky localisation region and that become more incomplete with redshift \citep{mastrogiovanni2022cosmology}. This requires an alternative and efficient approach to tracking host galaxies with GW events.  
\begin{table*}
    \centering
    \begin{tabular}{ccccccccc}
        \hline
        \hline
        \multirow{3}{*}{Survey} &  \multirow{3}{*}{Redshift} & Survey & \multirow{3}{*}{Target} & {\sc Shark}  & \multicolumn{4}{|c|}{Observed merger rate density} \\
        & & Area &  & density &  \multicolumn{4}{|c|}{(Myr$^{-1}$/$\rm deg^2$)} \\ 
        & & ($\rm deg^2$) & (gal/$\rm deg^2$) & (gal/$\rm deg^2$) & Fryer & Mandel \& Muller & Mandel \& Muller & Schneider \\
        & & & & & & ($f_{\rm WR}=1$) & ($f_{\rm WR}=0.2$) & \\
        \hline
        4MOST CRS & $0.15<z<0.4$ & 7500 & 250 & 250 & $43_{-5}^{+5}\times 10^2$ & $48_{-5}^{+5}\times 10^1$ & $66_{-7}^{+7}\times 10^1$ & $53_{-6}^{+6}\times 10^0$ \vspace{1mm} \\
        4MOST HS & $z<0.15$ & 20,000& 325 & 325 & $70_{-3}^{+5}\times 10^2$ & $10_{-1}^{+1}\times 10^2$ & $12_{-1}^{+1}\times 10^2$ & $87_{-4}^{+7}\times 10^0$ \vspace{1mm} \\
        DESI BG & $0.15<z<0.4$ & 14,000 & 800 & 800 & $20_{-3}^{+1}\times 10^3$ & $31_{-5}^{+2}\times 10^2$ & $34_{-5}^{+2}\times 10^2$ & $24_{-4}^{+2}\times 10^1$ \vspace{1mm} \\
        DESI ELG & $0.6<z<1.6$ & 14,000 & 2400 & 2400 & $11_{-1}^{+1} \times 10^4$ & $21_{-2}^{+2} \times 10^3$ & $20_{-2}^{+2} \times 10^3$ & $11_{-1}^{+1}\times 10^2$ \vspace{1mm} \\
        DESI LRG & $0.3<z<0.8$ & 14,000 & 610 & 610 & $40_{- 9}^{+16} \times 10^3$ & $75_{-16}^{+29} \times 10^2$ & $79_{-17}^{+31} \times 10^2$ & $90_{-20}^{+35} \times 10^1$ \vspace{1mm} \\ 
        DESI LIS & $z<7$ &  14,756 & - & 11,151 & $28_{-4}^{+ 5} \times 10^4$ & $50_{-7}^{+9} \times 10^3$ & $59_{-8}^{+10} \times 10^3$ & $30_{-4}^{+5}\times 10^2$ \vspace{1mm} \\
        LSST WFD & $z<4$ & 18,000 & - & 108,738 & $29_{-2}^{+3} \times 10^5$ & $51_{-4}^{+5} \times 10^4$ & $51_{-4}^{+5} \times 10^4$ & $34_{-3}^{+3} \times 10^3$ \vspace{1mm} \\
        \hline
        Full simulation & $z<3$ & 1.7 & - & 643,068 & $54_{-4}^{+4} \times 10^5$ & $94_{-7}^{+8} \times 10^4$ & $93_{-7}^{+8} \times 10^4$ & $72_{-5}^{+6} \times 10^3$ \\
        \hline
    \end{tabular}
    \caption{Summary of survey area, target density in survey and {\sc Shark} and expected merger rate density from the galaxies observed in each survey. for \citet{fryer2012compact}, \citet{mandel2020simple} (with $f_{\rm WR}=1$ and $f_{\rm WR}=0.2$) and \citet{schneider2021pre} remnant mass models.}
    \label{tab:selections}
\end{table*}
\begin{figure*}
    \centering
    \includegraphics[width=\textwidth]{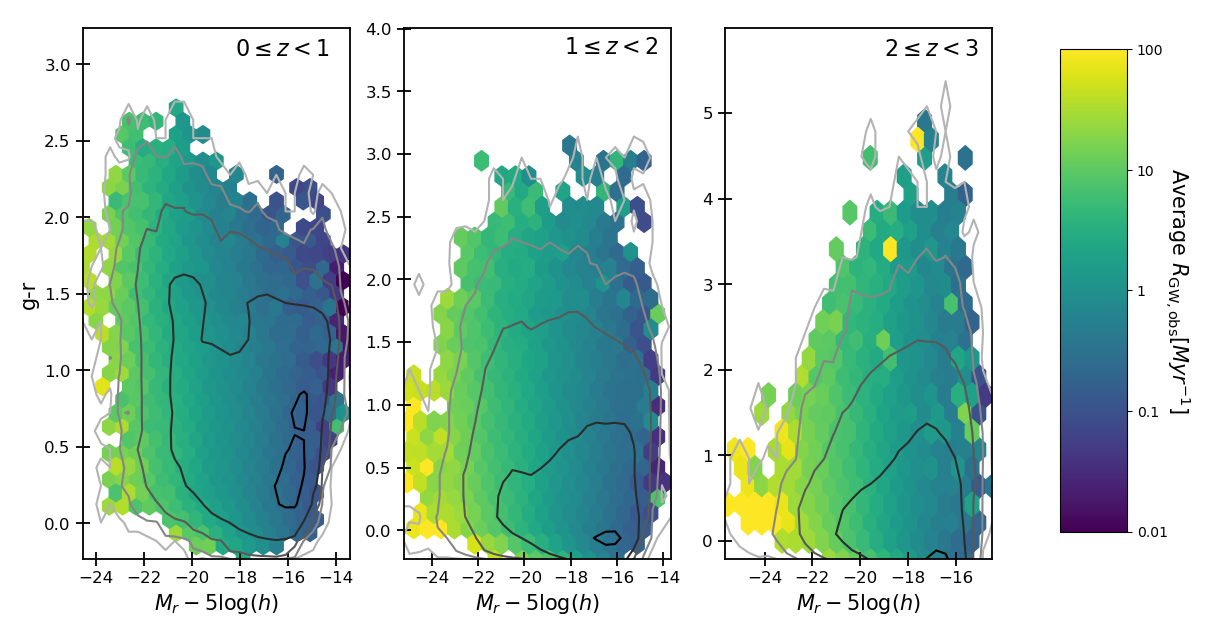}
    \caption{Colour-magnitude ($r$-band absolute magnitude) diagram of the host galaxies at $0 \leq z < 1$, $1 \leq z < 2$ and $2 \leq z < 3$. The colour code represents the average number of GW events per Myr in each bin. The contours represent the number of galaxies. }
    \label{fig:colour_mag}
\end{figure*}
The following list describes the different optical surveys we consider in this work, while the exact selection functions we implemented are given in Appendix~\ref{Cuts}: 
\begin{itemize}
    \item \textbf{4MOST Hemisphere Survey (HS):} A spectroscopic galaxy redshift survey, with a near-infrared selection (J-band) and a very high completeness over the southern hemisphere. The goals of this survey are to study environmental effects and processes, measure the effects of gravity and the growth rate of structure \citep{4mosths}. 
    \item \textbf{4MOST Cosmology Redshift Survey (CRS):} A spectroscopic survey targeting bright galaxies, luminous red galaxies, emission line galaxies and quasars. These probes will contribute to tests of gravity, analysing redshift distributions and enable synergies with other surveys \citep{richard20194most}. We chose the selections for the bright galaxy catalogue, as this would maximise the total merger rate (see Fig.~\ref{fig:colour_mag}) as it covers the largest area (although not necessarily the galaxies with the highest average GW event rate).
    \item \textbf{DESI Legacy Imaging Survey (LIS):} Consists of three photometric sub-surveys in the $grz$ optical bands and mid-infrared bands from WISE designed to target cosmological tracers. The sub-surveys are; The Dark Energy Camera Legacy Survey (DECaLS), Beijing-Arizona Sky Survey (BASS) and Mayall z-band Legacy Survey (MZaLS) \citet{2019AJ....157..168D}. We take the brightest magnitudes given for $3+$ exposures across the three surveys, to ensure our predictions are robust but conservative for the entire survey area of $14,000$ deg$^2$.
    \item \textbf{DESI Bright Galaxy Survey (BGS):} This $r$-band magnitude limited spectroscopic survey, consisting of a bright and faint sample, is designed for clustering analyses yielding precise measurements of baryonic acoustic oscillations and redshift space distortions \citep{2021MNRAS.tmp..319R}. We chose the bright sample selection, which has a 95\% redshift completeness and is expected to maximise our GW event rate based on Fig.~\ref{fig:colour_mag}. 
    \item \textbf{DESI Luminous Red Galaxy (LRG) Sample:} This survey will obtain spectra for more than 8 million luminous red galaxies. LRGs are massive galaxies which have ceased star formation. However, due to the $4000\mathring{A}$ break feature, the redshifts are easily obtained. LRGs are ideal tracers of larger scale structure in the Universe. The LRG targets are selected from the Legacy Imaging surveys (see below; \citealt{zhou2020preliminary}). We find that using the colour and magnitude cuts in the northern (BASS/MZaLS) sample give us a predicted density closest to the target density.
    \item \textbf{DESI Emission Line Galaxy (ELG) Sample:} This spectroscopic survey will make use of the abundance of distant but highly star-forming Emission Line Galaxies (ELGs). The selection consists of a $g$-band magnitude cut and a ($g-r$) vs ($r-z$) colour cut \citep{raichoor2020preliminary}. We use the selections for the south region, as this gives us a {\sc Shark} predicted galaxy density closest to the target density quoted in \cite{raichoor2020preliminary}. 
    \item \textbf{Legacy Survey of Space and Time (LSST):} A planned six band (\textit{ugrizy}) photometry survey with a duration of 10 years on the Simonyi Survey Telescope at the Vera C. Rubin Observatory. It aims to, among other things, catalog a transformative number of transients, study galaxy formation and properties, and to constrain the nature of dark energy and dark matter \citep{abell2009lsst}. We chose the median co-added $5\sigma$ depths in the Wide-Fast-Deep area. Out of the 6 bands, we also chose the $grz$-band cuts so we could compare this to the Legacy Imaging Survey. 
\end{itemize}
In implementing the selection for the above surveys, all spectroscopic survey cuts were altered up to 0.3 mag so that the galaxy density obtained in the simulation is close to the target density of the survey as quoted in the relevant reference (see Appendix~\ref{Cuts} for further details). This is reasonable as some of the galaxies may be insufficiently resolved in the simulation and a few tenths of a mag uncertainty is expected when producing the predicted galaxy SEDs from the SFR and metallicity histories \citep{2019MNRAS.489.4196L}. In all cases we are able to accurately recreate the predicted redshift distributions for each survey as shown in Figure~\ref{fig:z_survey_dist}. 

\begin{figure*}
    \centering
    \includegraphics[width=\textwidth]{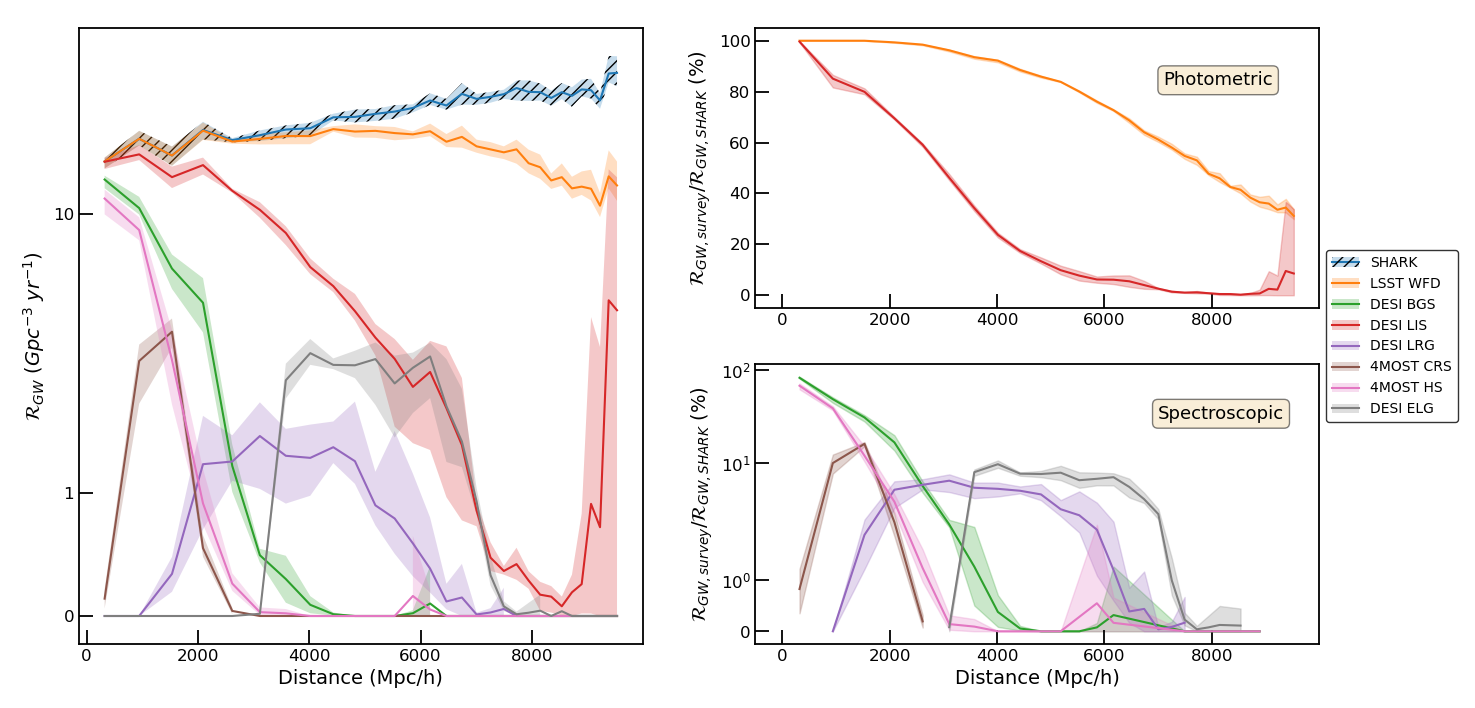}
    \caption{Volumetric rate of GWs emitted from galaxies observed by different surveys, as a function of distance. The lines show the average volumetric rate over the subvolumes, while the shaded regions are the range. The right plots show the effective completeness, which is the fraction of GW host galaxy mergers observed by each survey, relative to the total number of expected mergers in the {\sc Shark} simulation volume. The large uncertainties in the plots for $>5000$~Mpc/h are due to small number statistics.}
    \label{fig:completeness}
\end{figure*}

Table \ref{tab:selections} summarises the different selection effects we apply to our GW catalogue and the host galaxies and the resulting merger rates. 
The left plot in Fig.~\ref{fig:completeness} shows the BBH  volumetric rate given the selection effects from the seven optical surveys described above. The right plot shows the merger rate completeness with respect to the {\sc Shark} merger rate. The photometric surveys do well to be essentially complete out to $2000$ $h^{-1}$ Mpc, with the deeper LSST imaging performing better at larger distances as one would expect. The completeness for DESI BGS and 4MOST HS is relatively high around $z=0$, but there is a steeper decline compared to LSST and DESI LIS. This is due to 80\% of the GW host galaxies being found at $z \geq 1$, which is not probed by these two magnitude limited surveys. 

To expand on this, we find it revealing to plot the GW rate as a function of absolute $r$-band magnitude and $g-r$ colour (as was investigated by \citealt{artale2019host}), in Fig.~\ref{fig:colour_mag}. Given the correlation between $r$ magnitude and stellar mass, we find that brighter, more massive galaxies are more likely to host a higher number of GW events out to $z=2$, regardless of the colour. In the $z=2-3$ range, the colour for galaxies with high merger rates can be constrained to $0 < g-r < 1.5$. Hence, any approximately magnitude limited survey, even the $J$-band and $J-K$ cuts for 4MOST HS that similarly target the bright galaxies, will do generally well to obtain the most prolific hosts across the entire volume they probe, but rapidly lose the larger number of more moderately contributing hosts as they fall out of the magnitude limit.


The DESI ELG, LRG and 4MOST CRS surveys are designed to sample galaxies in a higher redshift range compared to 4MOST HS.  In all cases, the maximum completeness they obtain is around $10\%$ and is limited by certain colour cuts. We find that the $(r-z)$ vs $(g-r)$ cut for DESI-ELGs removes a large portion of the GW host galaxies with high merger rates, while the DESI-LRG $r-W1$ vs $W1$ cut also excludes a majority of the high-rate GW host galaxies particularly those around $W1 \sim 20$ in the $1 \leq z < 2$ range. One interesting feature however is that these surveys find a higher merger rate density than the lower redshift surveys even accounting for the larger galaxy density. This is by virtue of Fig.~\ref{fig:RGWvol}, where the merger rate density increases up to $z \sim 2$ such that there are more BBH mergers in the higher redshift redshift range to capture. 


\begin{figure*}
    \centering
    \includegraphics[width=\textwidth]{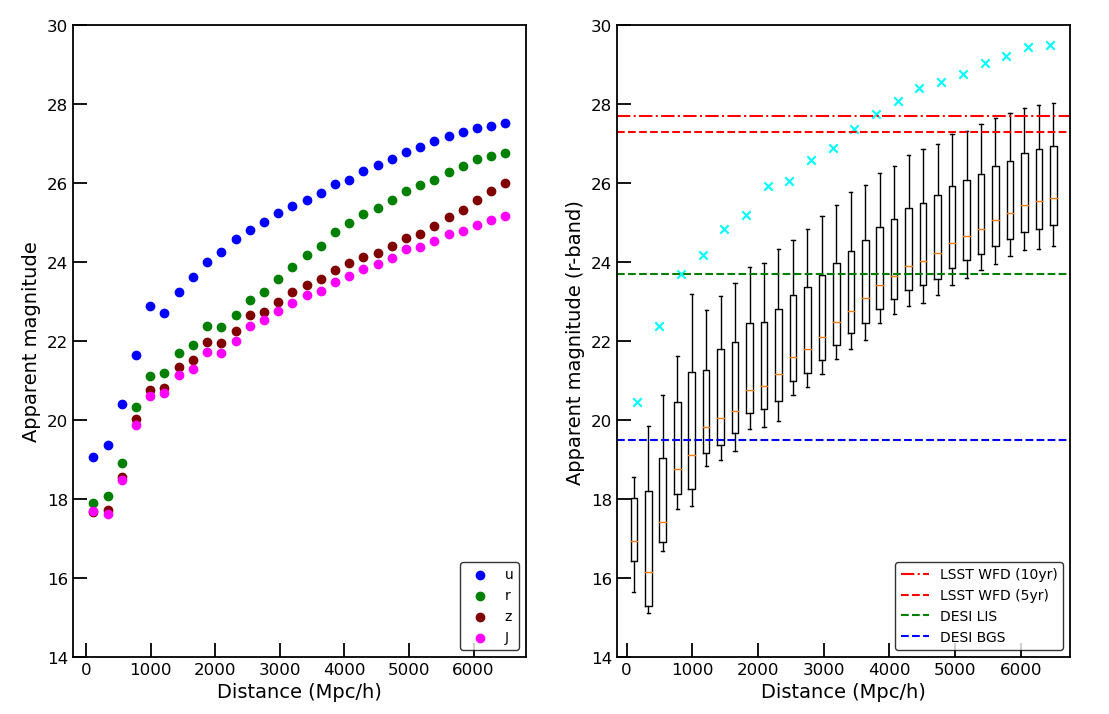}
    \caption{Host galaxy merger rate completeness as a function of distance. The left plot shows the magnitude limit for galaxies that contain 75\% of the total merger rate in each distance bin. The right plot shows the $r$-band apparent magnitude for galaxies at 25\%, 35\%, 50\%, 75\% and 90\% completeness. For comparison, we also include the r-band selections for LSST at the 5 year \citep{ivezic2019lsst} and 10 year \citep{abell2009lsst} mark, for DESI BGS and DESI LIS. The cyan crosses show the 90th percentile for all the galaxies in the simulation (regardless of whether or not they host a GW).} 
    \label{fig:box_plot}
\end{figure*}
In Fig.~\ref{fig:box_plot}, we flip our investigation around and consider what kinds of surveys would be required to capture the majority GW hosts within a given distance. The left plot of this figure shows the magnitude cut required to contain $75\%$ of the total merger rate in each distance bin. The $u$ and $r$-band go deeper than the $z$ and $J$-band out to further distances. In the right plot, we further analyse the merger rate completeness in the $r$-band. This band is common to most photometric and spectroscopic surveys and correlates strongly with the merger rate. DESI BGS does well to observe galaxies with at least $50\%$ of the total merger rate out to $1000$ $h^{-1}$ Mpc. DESI LIS is slightly better, as it will observe host galaxies containing $90\%$ of the total merger rate out to $2000$ $h^{-1}$ Mpc. However, LSST sets the precedent for the required $r$-band depth, since it is able to observe host galaxies containing $90\%$ of the total merger rate out to $6500$ $h^{-1}$ Mpc. Note that the cyan crosses are included to show that $90\%$ of the total merger rate is contained within $<90\%$ of the galaxies in each redshift bin. Therefore, these brighter galaxies must be prioritised when searching for GW events, as they will have higher merger rates. If we choose LSST as our ideal survey for observing the host galaxies of BBH mergers, using the merger rate density and survey area we would expect up to $\sim$910 BBH mergers with host galaxies available from LSST to occur per year. The actual detection rate will depend strongly on the capabilities of GW detectors. This seems reasonable as LIGO/Virgo have detected 82 BBH mergers in the past five years, and we expect the detection rate to be much smaller than the actual merger rate.\par 
The marginal primary mass versus redshift distribution in GWTC3 shows the maximum redshift measured to be around 1.4, with majority of detections clustering around 0.3 \citep{ligo2021population}. In this case, having a survey like 4MOST CRS, DESI BG and DESI LRG would be ideal for tracking these events. However, given that GWTC-2 only detected GW sources out to $z<1$, it is clear the GW redshift distribution will be extended with future detections. In Fig.~\ref{fig:RGWvol}, we find the peak volumetric rate around $z=2.53$, so DESI LIS and LSST-like surveys are more ideal for tracing these high redshift galaxies. As these are photometric surveys, efficient spectroscopic follow-up will be crucial to give us a better picture of the GW rate versus redshift distribution for future applications in cosmology. A larger catalogue of GW events will also give us a better picture of the BH mass and spin distribution, determine if these properties evolve with redshift and possibly reveal more about the origins of BBHs \citep{kalogera2019deeper}. 

For each survey we have considered, we are mainly interested in the completeness \textit{fraction}, hence anything that affects the GW merger rate universally, such as a change in the COMPAS fraction of Section~\ref{sec:feff}, will leave the completeness fraction unchanged, as this is a relative fraction. However, as with the rest of this work, there are caveats that the exact metallicity distribution, galaxy star formation rates and COMPAS input parameters may change the predictions for a given survey. As discussed earlier, the impact of the delay time has a significant impact in the local Universe. This is also evident in Table \ref{tab:selections}, where the predicted volumetric rate in each survey varies depending on the remnant mass model. 

\section{Conclusion} \label{Section 7} 
In this work, we have produced and analysed a comprehensive simulation of GW events from BBHs and their host galaxies. By combining the semi-analytic model {\sc Shark} with the population synthesis code COMPAS we have estimated the BBH merger rate as a function of time and also determined relationships between the merger rate and host galaxy properties. 
In Section \ref{Section 2}, we justify the use of {\sc Shark} for our merger rate modelling and how the SFR and metallicity distribution are within the uncertainty range of the observational data. In Section \ref{Section 3} we discuss how we extract the BBH information required and perform many realisations to obtain the uncertainty from the stellar evolution models. We provide a detailed method of the computational process for modelling the BBH merger rate in Section \ref{Section 4}. \par 
We use the observed redshifts from the lightcone to determine the observed merger rate for each galaxy and find the relationship between their host galaxy properties in Section \ref{Section 5}. We found a strong correlation between the merger rate and stellar mass, which is a consequence of the relationship between SFR and stellar mass. We also compare the merger rate density to the observational fit from GWTC-3. The discrepancy between our default case \citet{fryer2012compact} and the observations most likely originates from the low redshift SFR overestimation, metallicity distribution, and hence the coalescence times. We decided to explore alternate remnant models and found that using the \citet{mandel2020simple} and \citet{mandel2020simple} with $f_{\rm WR}=0.2$ models would align our merger rate density closer to the GWTC-3 prediction. We also provide fitting formulae for the GW rate as a function of host galaxy properties and redshift, as well as a fitting formula for the total volumetric rate versus redshift. We find that both the \citet{madau2014cosmic} and right skew-normal distribution fitting formula best matches our model. \par  
In section \ref{Section 6}, our analysis aimed to answer two questions: `What is the merger rate completeness for future optical surveys, as a function of redshift?' \textit{and} `What is the apparent magnitude required for 25\%, 35\%, 50\%, 75\% and 90\% completeness, at fixed distances?'. We find that LSST is ideal for tracing galaxies with high merger rates, which is strongly dependent on the $r$-band magnitude. Given the $r$-band magnitude cut for LSST (10th year), we would be able to trace at least 90\% of the total number of mergers out to $6000$ $h^{-1}$ Mpc. \par
In this paper, we have considered BBHs formed via the isolated channel. Including other formations channels would strengthen the constraints on the merger rates and require us to incorporate star or globular clusters in our simulations. However, we have shown the benefits of using simulations to model BBH mergers. It enables us to make predictions at all redshifts and gives us insight into the distribution of BBH properties, to constrain stellar evolution models. In future work, we aim to use this simulation to investigate how the various incompleteness of the surveys we have considered may affect cosmological constraints from GWs and be overcome.

\section*{Acknowledgements}
We sincerely thank Ilya Mandel for his suggestions on aspects of this work and insight into which COMPAS input parameters to explore for this paper. We thank Jeff Riley and Mike Lau for assisting in the initial set up of COMPAS and with interpreting the compiler files. We also thank Matías Bravo for helpful discussions regarding {\sc Shark}.
CL has received funding from the Australian Research Council Centre of
Excellence for All Sky Astrophysics in 3 Dimensions (ASTRO 3D), through project number CE170100013. 
This work was supported by resources provided by The Pawsey Supercomputing Centre with funding from the 
Australian Government and the Government of Western Australia.  This research was supported by the Australian Government through the Australian Research Council Laureate Fellowship grant FL180100168.

We thank Christopher Berry for providing feedback on the first version of the paper. We also thank the reviewer for taking the time and effort to review the manuscript. We appreciate their valuable comments, which helped improve the manuscript. 
\section*{Data Availability}
A subset of the merger rate histories or observed merger rates for the STINGRAY lightcone galaxies can be provided upon request. Simulations in this paper made use of the COMPAS rapid binary population synthesis code (version v02.19.04), which is freely available at  \href{http://github.com/TeamCOMPAS/COMPAS}{this repository}.  

The SURFS simulation used in this work can be accessed freely from https://tinyurl.com/y6ql46d4. {\sc Shark} and STINGRAY are public codes (see Section~\ref{Section 2}).



\bibliographystyle{mnras}
\bibliography{example} 




\appendix

\section{Optical survey cuts} \label{Cuts}



Table~\ref{tab:selectioncuts} details the various cuts we apply to our simulated GW and host catalogue to reproduce the number density of galaxies and their redshift distribution for each survey we consider in this work and in turn predict the GW host completeness fraction.

\begin{table*}
    \centering
    \begin{tabular}{c|c|c}
        Survey Name & Actual (AB mag) & Adjusted (AB mag) \\
        \hline
        \hline
        \multirow{3}{*}{4MOST CRS BG} & 16 < J < 18.25 & 16 < J < 18.298 \vspace{2mm} \\ 
        & J $-$ W1 > 1.6 $\times$ (J $-$ K) $-$ 1.6 AND & 1.6 $\times$ (J $-$ K) $-$ 1.4 \\
        & J $-$ W1 < 1.6 $\times$ (J $-$ K) $-$ 0.5 & J $-$ W1 < 1.6 $\times$ (J $-$ K) $-$ 0.4 \vspace{2mm} \\
        & J $-$ W1 > $-$2.5 $\times$ (J $-$ K) $+$ 0.1 AND & No changes \\
        & J $-$ W1 < $-$0.5$\times$ (J $-$ K) $+$ 0.1 & No changes \\ 
        \hline
        4MOST HS & J $-$ K < 0.45 & J $-$ K < 0.4 \\
        & J < 18 & J < 17.951 \\
        \hline
        DESI BGS & r < 19.5 & r < 19.5375 \\
        \hline
        \multirow{4}{*}{DESI ELG} & 20 < g < 23.5 & \multirow{4}{*}{No changes} \\
        & 0.3 < r $-$ z < 1.6 & \\
        & g $-$ r < 1.15 $\times$ (r $-$ z) $-$ 0.15 & \\
        & g $-$ r < $-$1.2 $\times$ (r $-$ z) $+$ 1.6 & \\
        \hline
        DESI LIS & g < 23.48, r < 22.87, z < 22.22 & No changes \\
        \hline
        \multirow{4}{*}{DESI LRG} & (z $-$ W1) > 0.8 $\times$ (r $-$ z) $-$ 0.6 & (z $-$ W1) > 0.8 $\times$ (r $-$ z) $-$ 0.4 \vspace{2mm}\\ 
        & z < 21.61 & z < 21.39 \vspace{2mm} \\ 
        & g $-$ W1 > 2.97 OR r $-$ W1 > 1.8  & g $-$ W1 > 3.17 OR r $-$ W1 > 2 \vspace{2mm}\\ 
        & r $-$ W1 > 1.83 $\times$ (W1 $-$ 17.13) AND  & 
        r $-$ W1 > 1.83 $\times$ (W1 $-$ 17.11) AND  \\ 
        & r $-$ W1 > (W1 $-$ 16.31) OR r $-$ W1 > 3.4 & r $-$ W1 > (W1 $-$ 16.09) OR r $-$ W1 > 3.4 \\
        \hline
        LSST WFD & g < 27.5, r < 27.7, z < 26.2 & No changes \\
        \hline 
    \end{tabular}
    \caption{Selection functions for different galaxy surveys considered in this work. See the main text for more details on each survey. The column `Actual' provides the formal selection function for each survey taken from its reference, whereas `Adjusted' provides the slightly modified cuts we apply to our simulation to reproduce the expected number density of galaxies and galaxy redshift distribution.}
    \label{tab:selectioncuts}
\end{table*}

\bsp	
\label{lastpage}
\end{document}